\documentclass[longauth]{aa} % for the long lists of affiliations 
\usepackage{graphicx,natbib}
\usepackage{txfonts}
\begin{document}
   \title{WEBT multiwavelength monitoring and XMM-Newton observations of \object{BL Lacertae} in 2007--2008} 

   \subtitle{Unveiling different emission components\thanks{The radio-to-optical 
   data presented in this paper are stored in the WEBT archive ({\tt http://www.oato.inaf.it/blazars/webt/}); 
   for questions regarding their availability,
   please contact the WEBT President Massimo Villata ({\tt villata@oato.inaf.it}).}}

   \author{C.~M.~Raiteri              \inst{ 1}
   \and   M.~Villata                  \inst{ 1}
   \and   A.~Capetti                  \inst{ 1}
   \and   M.~F.~Aller                 \inst{ 2}
   \and   U.~Bach                     \inst{ 3}
   \and   P.~Calcidese                \inst{ 4}
   \and   M.~A.~Gurwell               \inst{ 5}
   \and   V.~M.~Larionov              \inst{ 6,7}
   \and   J.~Ohlert                   \inst{ 8}
   \and   K.~Nilsson                  \inst{ 9}
   \and   A.~Strigachev               \inst{10}
   \and   I.~Agudo                    \inst{11}
   \and   H.~D.~Aller                 \inst{ 2}
   \and   R.~Bachev                   \inst{10}
   \and   E.~Ben\'{i}tez              \inst{12}
   \and   A.~Berdyugin                \inst{ 9}
   \and   M.~B\"ottcher               \inst{13}
   \and   C.~S.~Buemi                 \inst{14}
   \and   S.~Buttiglione              \inst{15}
   \and   D.~Carosati                 \inst{16}
   \and   P.~Charlot                  \inst{17,18}
   \and   W.~P.~Chen                  \inst{19}
   \and   D.~Dultzin                  \inst{12}
   \and   E.~Forn\'e                  \inst{20}
   \and   L.~Fuhrmann                 \inst{ 3}
   \and   J.~L.~G\'{o}mez             \inst{11}
   \and   A.~C.~Gupta                 \inst{21}
   \and   J.~Heidt                    \inst{22}
   \and   D.~Hiriart                  \inst{12}
   \and   W.-S.~Hsiao                 \inst{19}
   \and   M.~Jel{\'i}nek              \inst{23}
   \and   S.~G.~Jorstad               \inst{24}
   \and   G.~N.~Kimeridze             \inst{25}
   \and   T.~S.~Konstantinova         \inst{ 6}
   \and   E.~N.~Kopatskaya            \inst{ 6}
   \and   A.~Kostov                   \inst{10}
   \and   O.~M.~Kurtanidze            \inst{25}
   \and   A.~L\"ahteenm\"aki          \inst{26}
   \and   L.~Lanteri                  \inst{ 1}
   \and   L.~V.~Larionova             \inst{ 6}
   \and   P.~Leto                     \inst{27,14}
   \and   G.~Latev                    \inst{28}
   \and   J.-F.~Le~Campion            \inst{17,18}
   \and   C.-U.~Lee                   \inst{29}
   \and   R.~Ligustri                 \inst{30}
   \and   E.~Lindfors                 \inst{ 9}
   \and   A.~P.~Marscher              \inst{31}
   \and   B.~Mihov                    \inst{10}
   \and   M.~G.~Nikolashvili          \inst{25}
   \and   Y.~Nikolov                  \inst{10,28}
   \and   E.~Ovcharov                 \inst{28}
   \and   D.~Principe                 \inst{13}
   \and   T.~Pursimo                  \inst{32}
   \and   B.~Ragozzine                \inst{13}
   \and   R.~M.~Robb                  \inst{33}
   \and   J.~A.~Ros                   \inst{20}
   \and   A.~C.~Sadun                 \inst{34}
   \and   R.~Sagar                    \inst{21}
   \and   E.~Semkov                   \inst{10}
   \and   L.~A.~Sigua                 \inst{25}
   \and   R.~L.~Smart                 \inst{ 1}
   \and   M.~Sorcia                   \inst{12}
   \and   L.~O.~Takalo                \inst{ 9}
   \and   M.~Tornikoski               \inst{26}
   \and   C.~Trigilio                 \inst{14}
   \and   K.~Uckert                   \inst{13}
   \and   G.~Umana                    \inst{14}
   \and   A.~Valcheva                 \inst{10}
   \and   A.~Volvach                  \inst{35}
 }

   \offprints{C.~M.~Raiteri}

   \institute{
 % 1
          INAF, Osservatorio Astronomico di Torino, Italy                                                     
 %           2
   \and   Department of Astronomy, University of Michigan, MI, USA                                            
 %           3
   \and   Max-Planck-Institut f\"ur Radioastronomie, Bonn, Germany                                            
 %           4
   \and   Osservatorio Astronomico della Regione Autonoma Valle d'Aosta, Italy                                
 %           5
   \and   Harvard-Smithsonian Center for Astroph., Cambridge, MA, USA                                         
 %           6
   \and   Astron.\ Inst., St.-Petersburg State Univ., Russia                                                  
 %           7
   \and   Pulkovo Observatory, St.\ Petersburg, Russia                                                        
 %           8
   \and   Michael Adrian Observatory, Trebur, Germany                                                         
 %           9
   \and   Tuorla Observatory, Dept.\ of Physics and Astronomy, Univ.\ of Turku, Piikki\"{o}, Finland          
 %          10
   \and   Inst.\ of Astronomy, Bulgarian Academy of Sciences, Sofia, Bulgaria                                 
 %          11
   \and   Instituto de Astrof\'{i}sica de Andaluc\'{i}a (CSIC), Granada, Spain                                
 %          12
   \and   Instituto de Astronom\'ia, Universidad Nacional Aut\'onoma de M\'exico, Mexico                      
 %          13
   \and   Department of Physics and Astronomy, Ohio Univ., OH, USA                                            
 %          14
   \and   INAF, Osservatorio Astrofisico di Catania, Italy                                                    
 %          15
   \and   SISSA-ISAS, Trieste, Italy                                                                          
 %          16
   \and   Armenzano Astronomical Observatory, Italy                                                           
 %          17
   \and   Universit\'e de Bordeaux, Observatoire Aquitain des Sciences de l'Univers, Floirac, France          
 %          18
   \and   CNRS, Laboratoire d'Astrophysique de Bordeaux -- UMR 5804, Floirac, France                          
 %          19
   \and   Institute of Astronomy, National Central University, Taiwan                                         
 %          20
   \and   Agrupaci\'o Astron\`omica de Sabadell, Spain                                                        
 %          21
   \and   ARIES, Manora Peak, Nainital, India                                                                 
 %          22
   \and   ZAH, Landessternwarte Heidelberg, Heidelberg, Germany                                               
 %          23
   \and   Inst.\ de Astrof{\'i}sica de Andaluc{\'i}a, CSIC, Spain                                             
 %          24
   \and   Inst.\ for Astrophysical Research, Boston University, MA, USA                                       
 %          25
   \and   Abastumani Astrophysical Observatory, Georgia                                                       
 %          26
   \and   Mets\"ahovi Radio Obs., Helsinki Univ.\ of Technology, Finland                                      
 %          27
   \and   INAF, Istituto di Radioastronomia, Sezione di Noto, Italy                                           
 %          28
   \and   Sofia University, Bulgaria                                                                          
 %          29
   \and   Korea Astronomy and Space Science Institute, South Korea                                            
 %          30
   \and   Circolo Astrofili Talmassons, Italy                                                                 
 %          31
   \and   Institute for Astrophysical Research, Boston University, MA, USA                                    
 %          32
   \and   Nordic Optical Telescope, Santa Cruz de La Palma, Spain                                             
 %          33
   \and   Dept.\ of Physics and Astronomy, Univ.\ of Victoria, Victoria, Canada                               
 %          34
   \and   Dept.\ of Phys., Univ.\ of Colorado Denver, Denver, CO USA                                          
 %          35
   \and   Radio Astronomy Lab.\ of Crimean Astrophysical Observatory, Ukraine                                 
 }

   \date{}
 
  \abstract
  % context heading (optional)
  % {} leave it empty if necessary  
   {BL Lacertae is the prototype of the blazar subclass named after it.
Yet, it has occasionally shown a peculiar behaviour that has questioned a simple interpretation of its broad-band emission in terms of synchrotron plus synchrotron self-Compton (SSC) radiation.}
  % aims heading (mandatory)
   {In the 2007--2008 observing season we carried out a new multiwavelength
 campaign of the Whole Earth Blazar Telescope (WEBT) on BL Lacertae, involving
 three pointings by the XMM-Newton satellite in July and December 2007, and January 2008, 
to study its emission properties, particularly in the optical--X-ray energy range.} 
%These observations were complemented by
%optical spectroscopic monitoring at the 3.6 m Telescopio Nazionale Galileo (TNG)
%to investigate the elusive H$\alpha$ broad emission line.}
  % methods heading (mandatory)
   {The source was monitored in the optical-to-radio bands by 37 telescopes.
The brightness level was relatively low. Some episodes of very fast variability were detected in the optical bands. Flux changes had larger amplitude at the higher radio frequencies than at longer wavelengths.}
  % results heading (mandatory)
   {The X-ray spectra acquired by the EPIC instrument onboard XMM-Newton 
are well fitted by a power law with photon index $\Gamma \sim 2$ and photoelectric absorption exceeding the Galactic value. However, when taking into account the presence of a molecular cloud on the line of sight, the EPIC data are best fitted by a double power law, implying a concave X-ray spectrum.
The spectral energy distributions (SEDs) built with simultaneous radio-to-X-ray data at the epochs of the XMM-Newton observations suggest that the peak of the synchrotron emission lies in the near-IR band, and show a prominent UV excess, 
besides a slight soft-X-ray excess. 
A comparison with the SEDs corresponding to previous observations with X-ray satellites shows that the X-ray spectrum is very variable, since it can change from extremely steep to extremely hard, and can be more or less curved in intermediate states.
We ascribe the UV excess to thermal emission from the accretion disc, and the other broad-band spectral features to the presence of two synchrotron components, with their related SSC emission. 
We fit the thermal emission with a black body law and the non-thermal components by means of a helical jet model. The fit indicates a disc temperature $\ga 20000 \rm \, K$ and a luminosity $\ga 6 \times 10^{44} \rm erg \, s^{-1}$.}
  % conclusions heading (optional), leave it empty if necessary 
  {}


%\titlerunning{Multifrequency observations of 3C 454.3}

   \maketitle
%
%________________________________________________________________

\section{Introduction}

BL Lacertae is the prototype of one of the two blazar subclasses, the BL Lac objects,
the other subclass being represented by the flat-spectrum radio quasars (FSRQs).
Common features of blazars are:
i) extreme flux variability at all wavelengths, from radio to $\gamma$-ray frequencies, on a wide variety of time scales, from long-term (months, years) oscillations to intra-day variability (IDV); 
ii) high radio and optical polarization;
iii) brightness temperatures exceeding the Compton limit;
iv) superluminal motion of the radio components.
The commonly accepted paradigm foresees that their non-thermal emission comes from a plasma jet closely aligned with the line of sight. The jet is generated by a supermassive black hole fed by infall of matter from an accretion disc. 
The broad-band spectral energy distribution (SED) of a blazar, given in the common 
$\log (\nu F_\nu)$ versus $\log \nu$ representation, shows two wide bumps. 
The low-energy bump, which extends from the radio to the optical--UV (for some BL Lacs up to X-ray) frequencies, is ascribed to synchrotron radiation by relativistic electrons in the jet.
The high-energy bump, covering the X-ray to $\gamma$-ray energies, 
is likely due to inverse-Compton scattering of seed photons off the relativistic electrons. 
According to the synchrotron self Compton (SSC) model, the seed photons are the synchrotron photons themselves.
In contrast, the external Compton (EC) scenario foresees that seed photons may enter the jet either directly from the accretion disc \citep[e.g.][]{der92}, or reprocessed by the broad line region \citep[e.g.][]{sik94} or hot corona surrounding the disc \citep[e.g.][]{ghi09}. 
SSC models usually fairly explain the SEDs of the low-luminosity blazars, i.e.\ the BL Lac objects, while EC models are needed to fit the SEDs of the FSRQs. 
However, recent multiwavelength studies on a number of blazars, which included observations by the $\gamma$-ray satellite AGILE, have shown that multiple SSC and/or EC components are necessary to explain the observed high-energy fluxes \citep[see e.g.][]{che08,puc08,ver09,dam09,don09b}.

In addition to these two non-thermal jet components, the SEDs of quasar-type blazars sometimes show a ``blue bump" in between, which is thought to be the signature of the thermal radiation emitted from the accretion disc.
Indeed, the spectra of these objects usually display prominent broad emission lines, which are most likely produced by photoionization of the broad line region due to the disc radiation.

On the contrary, BL Lacs are by definition almost featureless objects (equivalent width less than 5 \AA\ in their rest frame, \citealt{sti91}).
It was hence a surprise when \citet{ver95}, and soon after \citet{cor96}, discovered a broad H$\alpha$ (and H$\beta$) emission line in the spectrum of BL Lacertae, 
whose luminosity ($\sim 10^{41} \rm \, erg \, s^{-1}$) and
full-width half-maximum ($\sim 4000 \rm \, km \, s^{-1}$) are comparable to those of type I Seyfert galaxies such as NGC 4151. 
Subsequent spectroscopic monitoring of this source by \citet{cor00} showed that 
the H$\alpha$ equivalent width is approximately inversely proportional to 
the optical continuum flux.
This suggested that the broad line region is photoionized by a radiation source
that is not the same producing the optical continuum.
The photoionising radiation would most likely come from the accretion disc.

Another important issue comes from the results obtained by \citet{rav03}. 
They analyzed the X-ray data acquired by BeppoSAX from October 31 to November 2, 2000, 
during an extensive multiwavelength campaign.
When constructing the source SED with contemporaneous data, it was evident that
the steep X-ray spectrum was offset with respect to the extrapolation of the optical one.
One possible explanation was the presence of an extra component in addition 
to the synchrotron and inverse-Compton ones.

Finally, according to \citet{mad99} and \citet{boe00}, the explanation of the $\gamma$-ray flux detected by the EGRET instrument on board the CGRO satellite during the 1997 optical outburst \citep[see][]{blo97} requires an EC emission component in addition to the SSC one.

Taken together, these results suggest that in BL Lacertae the interpretation of the broad-band emission may require a more complex scenario than that usually envisaged for the BL Lac objects, involving just one synchrotron plus its SSC emission.

In the last decade, BL Lacertae has been extensively studied by the Whole Earth Blazar Telescope (WEBT) collaboration\footnote{{\tt http://www.oato.inaf.it/blazars/webt/}},
which has carried out several multiwavelength campaigns on this object \citep{vil02,rav02,boe03,vil04a,vil04b,bac06,pap07,vil09}, collecting tens of thousands of optical-to-radio data.
These studies were focused on its multiwavelength flux variability, colour behaviour, correlations among flux variations in different bands, possible periodicity of the radio outbursts. 
The main aim of the new WEBT campaign organized in the 2007--2008 observing season was instead to address the problem of disentangling the possible multiple contributions to the BL Lac flux from the radio band to $\gamma$-rays. For this sake, the optical-to-radio monitoring by the WEBT was complemented by three pointings by the XMM-Newton satellite. Moreover, we also obtained optical spectra with the 3.6 m Telescopio Nazionale Galileo (TNG)
% in Spain, and at the 2.1 GHAO telescope in Mexico, 
to investigate the properties of the H$\alpha$ broad emission line, and possibly infer information on the accretion disc. The results of the spectroscopic study will be reported elsewhere.

This paper is organised as follows. 
In Sect.\ 2 we present the WEBT optical-to-radio light curves. 
The analysis of the XMM-Newton data is reported in Sect.\ 3. 
In Sect.\ 4 we show the SEDs corresponding to the XMM-Newton epochs, and compare them with those related to previous observations by X-ray satellites.
The interpretation of the XMM-Newton SEDs is discussed in Sect.\ 5.
Finally, Sect.\ 6 contains a summary and discussion of the main results.

\section{Multifrequency observations by the WEBT}

\begin{figure*}
   \sidecaption
   \includegraphics[width=12cm]{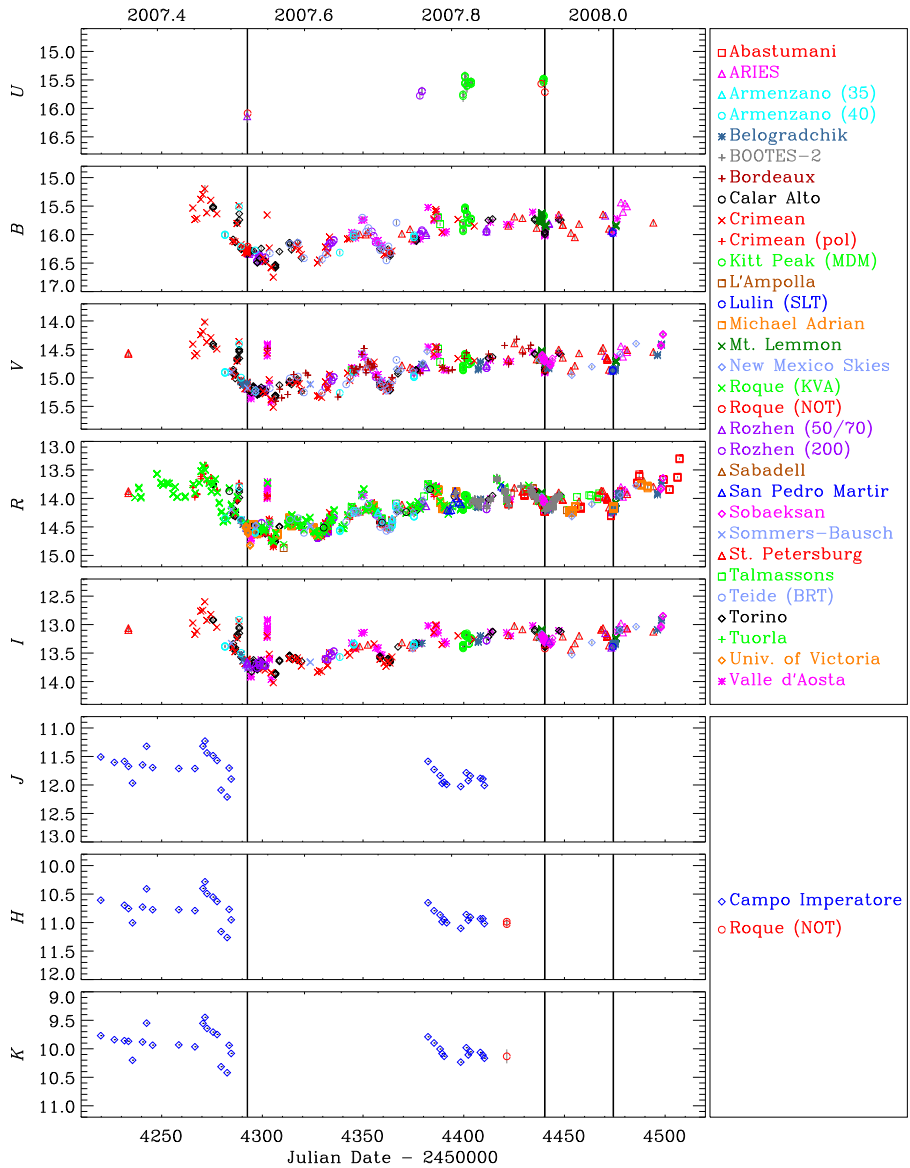}
      \caption{Optical $UBVRI$ and near-IR $JHK$ light curves of BL Lacertae in the 2007--2008 observing
 season. Vertical lines indicate the three XMM-Newton pointings of July 10--11 and December 5, 2007, and January 8, 2008.}
         \label{otir}
   \end{figure*}

The new WEBT campaign on BL Lacertae took place in the 2007--2008 observing season.
The participating observatories are listed in Table \ref{obs}.
Optical and near-IR data were collected as instrumental magnitudes of the source and reference stars in the same field to apply the same calibration \citep{ber69,fio96}.
The light curves obtained by assembling all datasets were carefully inspected to correct for systematic offsets and to reduce data scatter by binning noisy data taken by the same observer within a few minutes.
The results are shown in Fig.\ \ref{otir}, where the vertical lines indicate the epochs of the three XMM-Newton pointings.
We can see a noticeable flux variability, which progressively increases its amplitude going from the $I$ to the $B$ band. By considering only the period of common monitoring (before $\rm JD =2454500$), the maximum variability amplitude  (maximum $-$ minimum) is 1.42, 1.46, 1.50, and 1.55 mag in the $I$, $R$, $V$, and $B$ bands, respectively.
The near-IR time coverage is inferior to the optical one, but the near-IR data are important to add information to the SED (see Sect.\ 5).

\begin{table}
\caption{List of optical, near-IR, and radio observatories contributing data to this work.}
\label{obs}
\centering
\begin{tabular}{l r c  }
\hline\hline
Observatory    & Tel.\ size    & Bands\\%               & $N_{\rm obs}$\\
\hline
\multicolumn{3}{c}{\it Optical}\\
Abastumani, Georgia      &  70 cm         & $R$                \\%Long=+34.0125 deg, Lat=+44.7266 deg, Alt=650 m
ARIES, India             & 104 cm         & $BVRI$             \\
Armenzano, Italy         &  35 cm         & $BRI$              \\
Armenzano, Italy         &  40 cm         & $BVRI$             \\
Belogradchik, Bulgaria   &  60 cm         & $VRI$              \\
BOOTES-2, Spain          &  30 cm         & $R$                \\
Bordeaux, France         &  20 cm         & $V$                \\
Calar Alto, Spain$^a$    & 220 cm         & $R$                \\
Crimean, Ukraine         &  70 cm         & $BVRI$             \\
Kitt Peak (MDM), USA     & 130 cm         & $UBVRI$            \\
L'Ampolla, Spain         &  36 cm         & $R$                \\%Long=+0.670136 deg, Lat=+40.807 deg, Alt=75 m
Lulin (SLT), Taiwan      &  40 cm        & $R$                \\
Michael Adrian, Germany  & 120 cm        & $R$                \\%Long=+8.4114 deg, Lat=+49.9254, Alt=103 m
Mt.\ Lemmon, USA         & 100 cm        & $BVRI$             \\
New Mexico Skies, USA    &  30 cm        & $VRI$              \\
Roque (KVA), Spain       &  35 cm        & $R$                \\
Roque (NOT), Spain       & 256 cm        & $UBVRI$            \\
Rozhen, Bulgaria         & 50/70 cm      & $BVR$              \\%Long=+24.7439 deg, Lat=+41.6931, Alt=1759 m
Rozhen, Bulgaria         & 200 cm        & $BVRI$             \\%Long=+24.7439 deg, Lat=+41.6931, Alt=1759 m
Sabadell, Spain          &  50 cm        & $R$                \\
San Pedro Martir, Mexico &  84 cm        & $R$                \\
Sobaeksan, South Korea   &  61 cm        & $VRI$              \\  
Sommers-Bausch, USA      &  61 cm        & $VRI$              \\
St.\ Petersburg, Russia  &  40 cm        & $BVRI$             \\
Talmassons, Italy        &  35 cm        & $BVR$              \\
Teide (BRT), Spain       &  35 cm        & $BVR$              \\
Torino, Italy            & 105 cm        & $BVRI$             \\
Tuorla, Finland          & 103 cm        & $R$                \\
Univ.\ of Victoria, Canada & 50 cm       & $R$                \\
Valle d'Aosta, Italy     &  81 cm        & $BVRI$             \\%Long=+7.47833 deg, Lat=+45.7895 deg, Alt=1600 m
\hline
\multicolumn{3}{c}{\it Near-infrared}\\
Campo Imperatore, Italy  & 110 cm        & $JHK$          \\
Roque (NOT), Spain       & 256 cm        & $HK$          \\
\hline
\multicolumn{3}{c}{\it Radio}\\
Crimean (RT-22), Ukraine & 22 m          & 37 GHz          \\%lat=+44:23:52.6, long=2h 15m 55.1s
Mauna Kea (SMA), USA     &$8 \times 6$ m$^b$ & 230, 345 GHz      \\
Medicina, Italy          & 32 m          & 5, 8, 22 GHz        \\
Mets\"ahovi, Finland     & 14 m          & 37 GHz              \\
Noto, Italy              & 32 m          & 43 GHz              \\
UMRAO, USA               & 26 m          & 4.8, 8.0, 14.5 GHz   \\
\hline
\multicolumn{3}{l}{$^a$ Calar Alto data were acquired as part of the MAPCAT (Monitoring}\\
\multicolumn{3}{l}{AGN with Polarimetry at the Calar Alto Telescopes) project.}\\
\multicolumn{3}{l}{$^b$ Radio interferometer including 8 dishes of 6 m size.}
\end{tabular}
\end{table}

The majority of the variability episodes have a time scale of a few days, but we can also recognise a long-term increasing trend starting from $\rm JD \sim 2454300$, as well as a few very fast events. One of these involved a brightening of about 0.9 mag in 24 hours, from $R \sim 14.6$ on $\rm JD=2454301.5$  to $R \sim 13.7$ the night after, when observations at the Valle d'Aosta Observatory showed a source brightening of $\sim 0.3$ mag in less than 3 hours. This behaviour was confirmed by observations in the $V$ and $I$ bands, ruling out that this rapid flux increase was an artifact.
Similar fast variations are not uncommon in BL Lacertae.
When analysing the 13248 $R$-band data acquired by the WEBT \citep[][and this paper]{vil02,vil04a,vil04b,vil09} during 1500 nights over 
more than 15 years, though with inhomogeneous sampling, we can distinguish between two kinds of rapid flux variability.
\begin{itemize}
\item Fast and noticeable intraday variations: we consider variations $\geq 0.25$ mag with rate $\geq 0.1$ mag/hour. 
These were found in 25 nights out of 677 nights where the observing time coverage is $\geq 2.5$ hours, 
with the maximum amplitude episode involving a change $\Delta R = 0.52$ in about 3.9 hours.
\item Large interday variations: we consider variations $\geq 0.75$ mag in $\leq 36$ hours. 
These were observed 6 times, and the above mentioned episode (about 0.9 mag in 24 hours) is the most extreme one. 
\end{itemize}

\begin{figure*}
   \sidecaption
   \includegraphics[width=12cm]{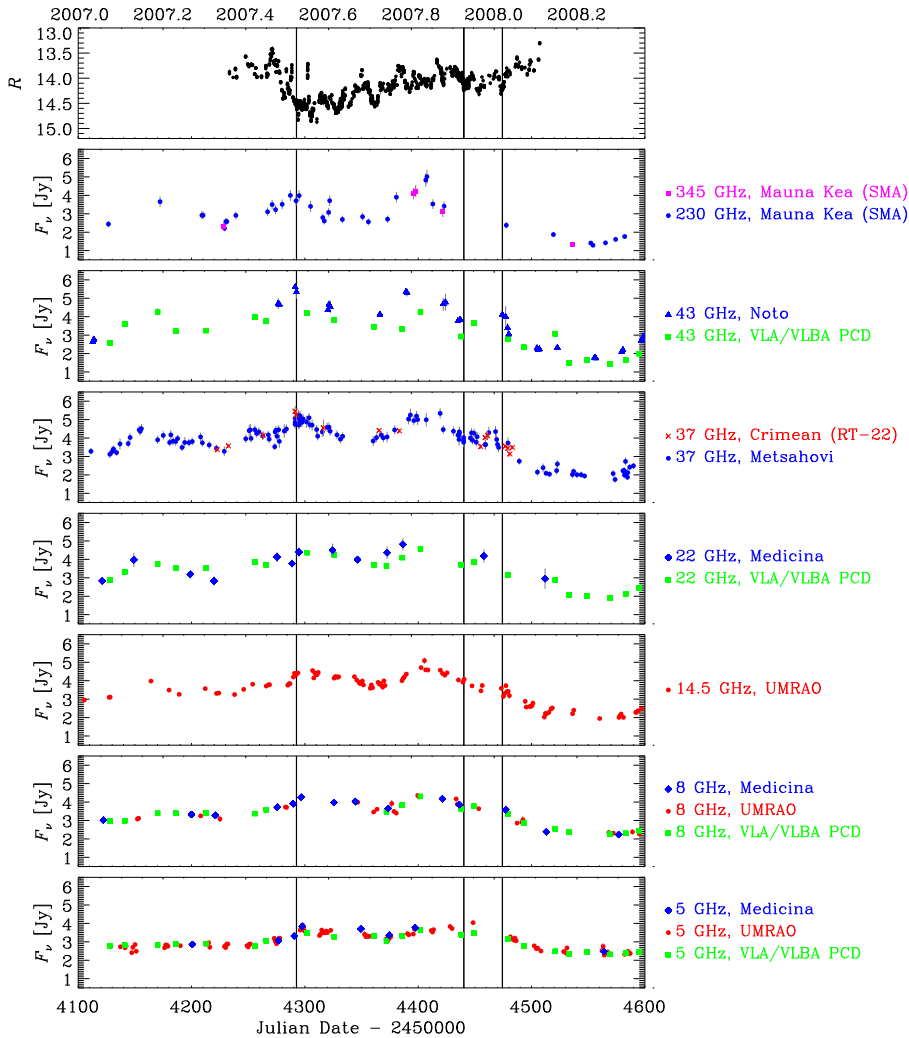}
      \caption{$R$-band light curve of BL Lacertae in 2007--2008 (top panel) compared to 
the behaviour of the radio flux densities (Jy) at different frequencies. 
Vertical lines indicate the three XMM-Newton pointings of July 10--11 and December 5, 2007, and January 8, 2008.}
         \label{radop}
   \end{figure*}

Radio data were collected as already calibrated flux densities. 
The radio light curves are shown in Fig.\ \ref{radop}, where the first panel displays the $R$-band light curve for comparison. We also included data from the VLA/VLBA Polarization Calibration Database (PCD)\footnote{\tt http://www.vla.nrao.edu/astro/calib/polar/}. As expected, the radio flux variations, which are more evident at the shortest wavelengths, are smoother than the optical variations\footnote{This is even more evident when comparing radio flux densities to optical flux densities instead of magnitudes.}, and the radio time scales are longer. Moreover, the long-term increasing trend characterising the optical light curve is not recognisable in the radio band.
According to \citet{vil09}, the optical outbursts of BL Lacertae are usually followed by high-frequency radio events, with time delays of at least 100 days, which can grow to 200 or even 300 days, depending on the relative orientation of the corresponding emitting regions in the jet. Hence, the high optical level observed at the beginning of our observing period, in May--June 2007, might be related to the bright radio state that is visible at the higher radio frequencies around JD = 2454400\footnote{Indeed, the PCD shows no further radio event until May 2009, the 43 GHz flux density remaining below 3 Jy.}.

\section{Observations by XMM-Newton}

The X-ray Multi-Mirror Mission (XMM) - Newton satellite observed the source on July 10--11 and December 5, 2007, and then on January 8, 2008 (PI: C.~M. Raiteri).

\subsection{EPIC data}

The European Photon Imaging Camera (EPIC) onboard XMM-Newton includes three detectors: 
MOS1, MOS2 \citep{tur01}, and pn \citep{str01}.
Since a bright state of the source could not be excluded, 
we chose a medium filter to avoid possible contamination
by lower-energy photons; moreover, we selected a small-window configuration
to minimize possible photon pile-up.

Data were reduced with the Science Analysis System (SAS) software, version 8.0.1,
following the same standard procedure adopted in \cite{rai07b}.
A temporal filtering was applied to discard high-background periods. 
%This produced a reduction of the available exposure time by about 20\% for the MOS detectors and by roughly 1/3 for the pn in both epochs.
Source spectra were extracted from circular regions of $\sim 40$ arcsec radius.
%The {\tt epatplot} task allowed us to check that pile-up was not affecting our observations
The MOS background was estimated on external CCDs, while for the pn we selected the largest source-free circle that could be arranged on the same CCD. 

The source spectra were binned  with the {\tt grppha} task of the FTOOL package, to have a minimum of 25 counts in each bin.
The binned spectra were then analysed in the 0.35--12 keV energy range with the {\tt Xspec} task of the XANADU package.
%, version 11.3.2.
We fitted the MOS1, MOS2, and pn spectra of each epoch together to increase the statistics.

\begin{figure}
   \resizebox{\hsize}{!}{\includegraphics[angle=-90]{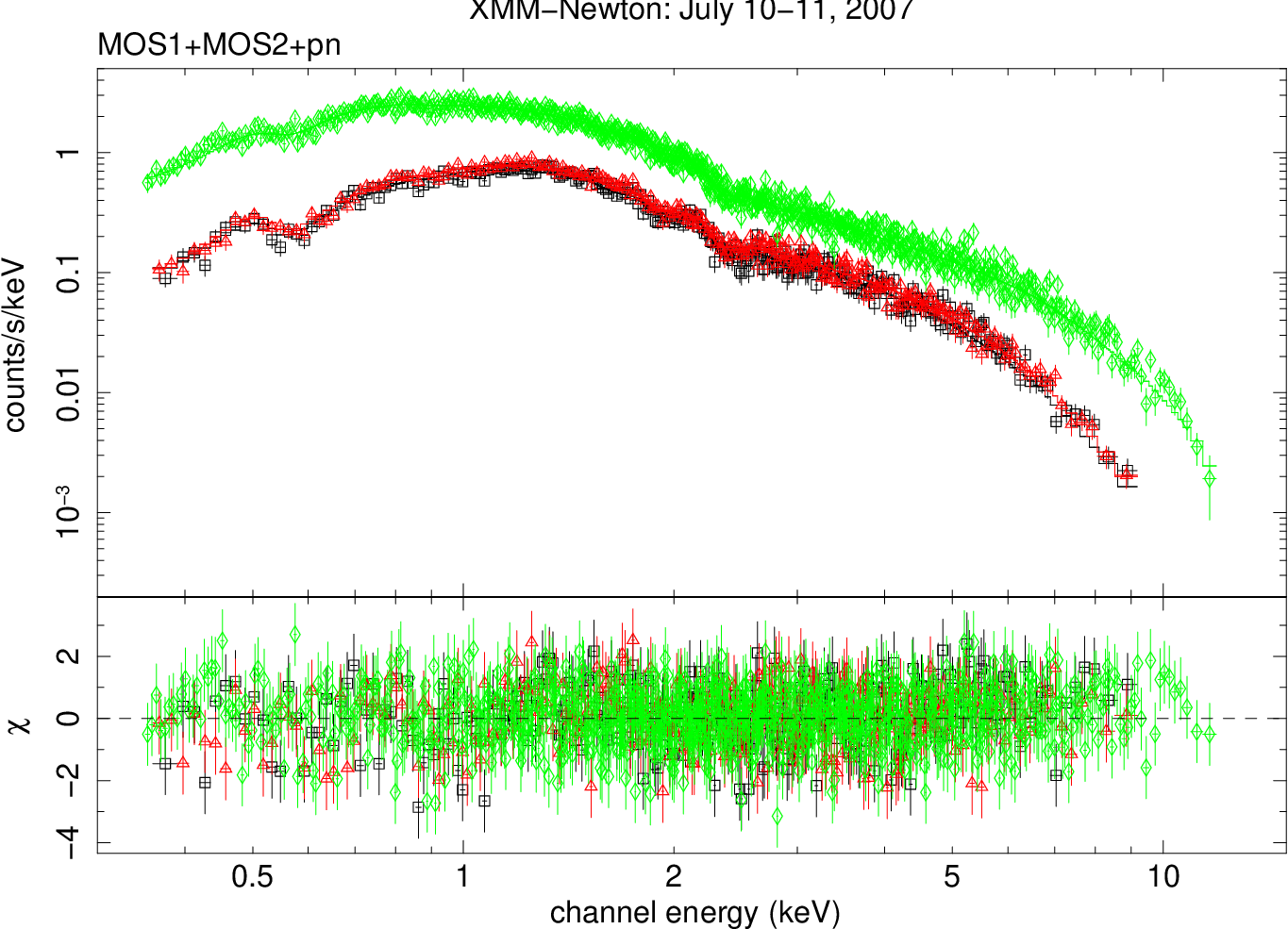}}
      \caption{EPIC spectrum of BL Lacertae on July 10--11, 2007; 
black squares, red triangles, and green diamonds represent MOS1, MOS2, and pn data, respectively.
The bottom panel shows the deviations of the observed data from the folded model (a power law with free absorption) in unit of standard deviations.
} 
         \label{xmm1}
   \end{figure}

\begin{figure}
   \resizebox{\hsize}{!}{\includegraphics[angle=-90]{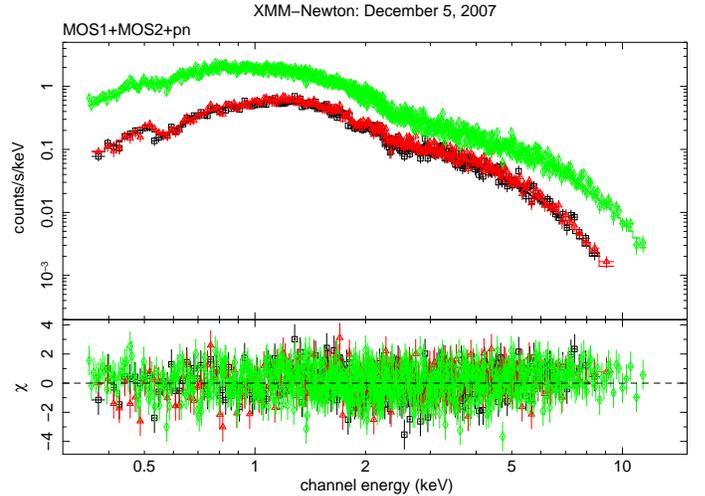}}
      \caption{EPIC spectrum of BL Lacertae on December 5, 2007; 
black squares, red triangles, and green diamonds represent MOS1, MOS2, and pn data, respectively.
The bottom panel shows the deviations of the observed data from the folded model (a power law with free absorption) in unit of standard deviations.
} 
         \label{xmm2}
   \end{figure}

\begin{figure}
   \resizebox{\hsize}{!}{\includegraphics[angle=-90]{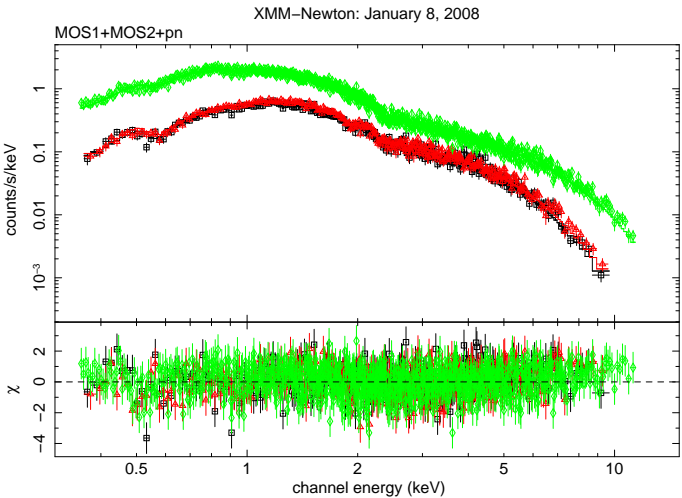}}
      \caption{EPIC spectrum of BL Lacertae on January 8, 2008; 
black squares, red triangles, and green diamonds represent MOS1, MOS2, and pn data, respectively.
The bottom panel shows the deviations of the observed data from the folded model (a power law with free absorption) in unit of standard deviations.
} 
         \label{xmm3}
   \end{figure}

We first considered a single power law model with free absorption, where the absorption is modelled according to the \citet{wil00} prescriptions.
The results of this spectral fitting for the three epochs are displayed in the top panels of 
Figs.~\ref{xmm1}--\ref{xmm3}. 
%the bottom panels show the residuals in terms of sigmas with error bars of size one. 
The corresponding model parameters are reported in Table~\ref{pow}, where 
Col.~2 gives the hydrogen column density,
Col.~3 the photon spectral index $\Gamma$,
Col.~4 the unabsorbed flux density at 1 keV,
Col.~5 the 2--10 keV observed flux,
and Col.~6 the value of $\chi^2/\nu$ (being $\nu$ the number of degrees of freedom ).
The $\chi^2/\nu$ values in Table~\ref{pow} indicate that the model is acceptable. This is also confirmed by the deviations of the observed data from the folded model, which are plotted in the bottom panels of Figs.~\ref{xmm1}--\ref{xmm3}.
The best-fit value of the hydrogen column density varies between 2.8 and $3.1 \times 10^{21} \, \rm cm^{-2}$, within the range of values found in previous analyses (see Sect.\ 5). 

Actually, the Galactic atomic hydrogen column density toward BL Lacertae 
is $N_{\rm H}=1.71 \times 10^{21} \, \rm cm^{-2}$ \citep[from the Leiden/Argentine/Bonn (LAB) Survey, see][]{kal05}. 
However, observations of local interstellar CO toward BL Lac have revealed a molecular cloud \citep{ban91,luc93,lis98}. According to \citet{lis98}, its $\rm ^{13}CO$ column density is $(8.48 \pm 0.78) \times 10^{14} \, \rm cm^{-2}$. Assuming that the molecular hydrogen column density $N_{\rm H_2}$ is typically $10^6$ times the $\rm ^{13}CO$ one \citep{luc93}, we derive an hydrogen column density of $\sim 1.7 \times 10^{21} \, \rm cm^{-2}$ due to the molecular cloud. This value depends on the uncertain ratio between CO and $\rm H_2$\footnote{From the recent paper by \citet{lis07} we can infer that $N_{\rm H_2} / N_{\rm ^{13}CO}$ is most likely in the range $\sim 1$--$2 \times 10^6$.}, but taking it at face value, the total hydrogen column density toward BL Lac becomes $N_{\rm H} = 3.4 \times 10^{21} \, \rm cm^{-2}$. This value is not very far from the $N_{\rm H}$ values we found when fitting the EPIC spectra with a power law model with free absorption, but as we will see in Sect.\ 5 this modest difference can make a difference in the interpretation of the source X-ray spectrum.

We thus fixed $N_{\rm H} = 3.4 \times 10^{21} \, \rm cm^{-2}$ and re-fitted a single power law model to the EPIC spectra. As expected, the goodness of the new fits is inferior to the previous case, and a slight excess of counts in the soft X-ray domain appears. This could be the signature of a curvature in the source spectra. Indeed, when we adopt a double power law model, the fit improves significantly, as is shown in Table \ref{double}, and this suggests that the spectrum is concave.

The $\chi^2/\nu$ values in Table~\ref{double} are a bit smaller than those in Table~\ref{pow}.
To better compare the two model fits, we calculated the F-test probability, which is $1.25 \times 10^{-2}$ for July 10--11, $2.70 \times 10^{-5}$ for December 5, and $3.11 \times 10^{-10}$ for January 8, 2008.
These results suggest that the double power law model with fixed absorption may be more appropriate to describe the EPIC spectra than the single power law model with free absorption. 

\begin{table*}
\caption{Results of fitting the EPIC data with a single power law with free absorption.}             
\label{pow}      
\centering  
\begin{tabular}{ c c c c c c}  
\hline\hline            
Date & $N_{\rm H}$ & $\Gamma$ & $F_{\rm 1 \, keV}$ & $F_{\rm 2-10 \, keV}$  & $\chi^2/\nu$ ($\nu$) \\
     & [$10^{21} \, \rm cm^{-2}$]&   & [$\mu$Jy]       & [$\rm erg \, cm^{-2} \, s^{-1}$]  &   \\     
 (1) & (2) & (3) & (4) & (5) & (6) \\
\hline                         
July 10--11, 2007  & 3.05 $\pm$ 0.06 & 2.01 $\pm$ 0.02 & 2.58 $\pm$ 0.04 & $9.64 \times 10^{-12}$ & 0.942 (1347)\\
December 5, 2007   & 2.92 $\pm$ 0.06 & 1.99 $\pm$ 0.02 & 1.96 $\pm$ 0.04 & $7.65 \times 10^{-12}$ & 1.026 (1250)\\
January 8, 2008    & 2.86 $\pm$ 0.06 & 1.91 $\pm$ 0.01 & 1.95 $\pm$ 0.03 & $8.51 \times 10^{-12}$ & 0.905 (1381)\\
\hline
\end{tabular}
\end{table*}

\begin{table*}
\caption{Results of fitting the EPIC data with a double power law with fixed total atomic and molecular column density. The two power laws have photon indices $\Gamma^1$ and $\Gamma^2$ and unabsorbed flux densities at 1 keV $F^1_{\rm 1 \, keV}$ and $F^2_{\rm 1 \, keV}$, respectively} 
\label{double}      
\centering  
\begin{tabular}{c c c c c c c c}  
\hline\hline            
Date & $N_{\rm H}$ & $\Gamma^1$ & $\Gamma^2$ & $F^1_{\rm 1 \, keV}$ & $F^2_{\rm 1 \, keV}$ & $F_{\rm 2-10 \, keV}$  & $\chi^2/\nu$ ($\nu$) \\
 & [$10^{21} \, \rm cm^{-2}$]&  &            & [$\mu$Jy]            & [$\mu$Jy]            & [$\rm erg \, cm^{-2} \, s^{-1}$] &   \\     
 (1) & (2) & (3) & (4) & (5) & (6) \\
\hline                                   
July 10--11, 2007  & 3.40 & 2.48 & 1.72 & 1.63 & 1.14 & $9.76 \times 10^{-12}$ & 0.939 (1346)\\
December 5, 2007   & 3.40 & 2.58 & 1.67 & 1.26 & 0.90 & $7.77 \times 10^{-12}$ & 1.013 (1249)\\
January 8, 2008    & 3.40 & 2.48 & 1.51 & 1.45 & 0.73 & $8.69 \times 10^{-12}$ & 0.880 (1380)\\
\hline
\end{tabular}
\end{table*}

\subsection{OM data}

XMM-Newton also carries an optical--UV 30 cm telescope, the Optical Monitor \citep[OM,][]{mas01}.
The BL Lac observations were performed with all its filters: $V$, $B$, $U$, UV$W1$, UV$M2$, and UV$W2$, with long exposures (see Table \ref{om}).
The OM data were reduced with the SAS software, version 8.0.1. 
The tasks {\tt omsource} and {\tt omphotom} were used to perform aperture photometry on the images produced by {\tt omichain}.
The resulting magnitudes are reported in Table \ref{om}. The uncertainties take into account the measure, systematic and calibration errors. In the optical filters, where a comparison with ground-based measurements is possible,
the OM magnitudes of the reference stars (B C H K) are within 0.1 mag with respect to the values we adopted for the calibration of the ground data, but they are stable (within 2-3 hundredths of mag) in the three XMM-Newton epochs.

\begin{table*}
\caption{Optical--UV magnitudes of BL Lacertae derived from the data analysis of the OM frames.} 
\label{om}      
\centering  
\begin{tabular}{c c c c c c c}  
\hline\hline            
Date & $V$ & $B$ & $U$ & UV$W1$ & UV$M2$ & UV$W2$ \\     
\hline                                   
\multicolumn{7}{c}{Exposure times (s)}\\
July 10--11, 2007 &  1498 & 1499 & 1499 & 2601 & 3098 & 6257$^a$ \\
December 5, 2007  &  2100 & 2099 & 2099 & 3500 & 3780 & 4000 \\
January 8, 2008   &  1700 & 1700 & 1698 & 2799 & 3300 & 7359$^b$   \\
\hline
\multicolumn{7}{c}{Magnitudes}\\
July 10--11, 2007 &  $15.33 \pm 0.10$ & $16.27 \pm 0.10$ & $15.87 \pm 0.10$ & $15.99 \pm 0.10$ & $16.90 \pm 0.10$ & $17.13 \pm 0.14$\\
December 5, 2007  &  $14.96 \pm 0.10$ & $15.91 \pm 0.10$ & $15.52 \pm 0.10$ & $15.62 \pm 0.10$ & $16.45 \pm 0.10$ & $16.98 \pm 0.11$\\
January 8, 2008   &  $14.97 \pm 0.10$ & $15.90 \pm 0.10$ & $15.53 \pm 0.10$ & $15.65 \pm 0.10$ & $16.47 \pm 0.10$ & $16.88 \pm 0.14$\\
\hline
\multicolumn{7}{l}{$^a$ Two exposures of 3579 and 2678 s.}\\
\multicolumn{7}{l}{$^b$ Two exposures of 3779 and 3580 s.}\\
\end{tabular}
\end{table*}

\section{Spectral energy distributions}

   \begin{figure*}
   \sidecaption
   \includegraphics[width=12cm]{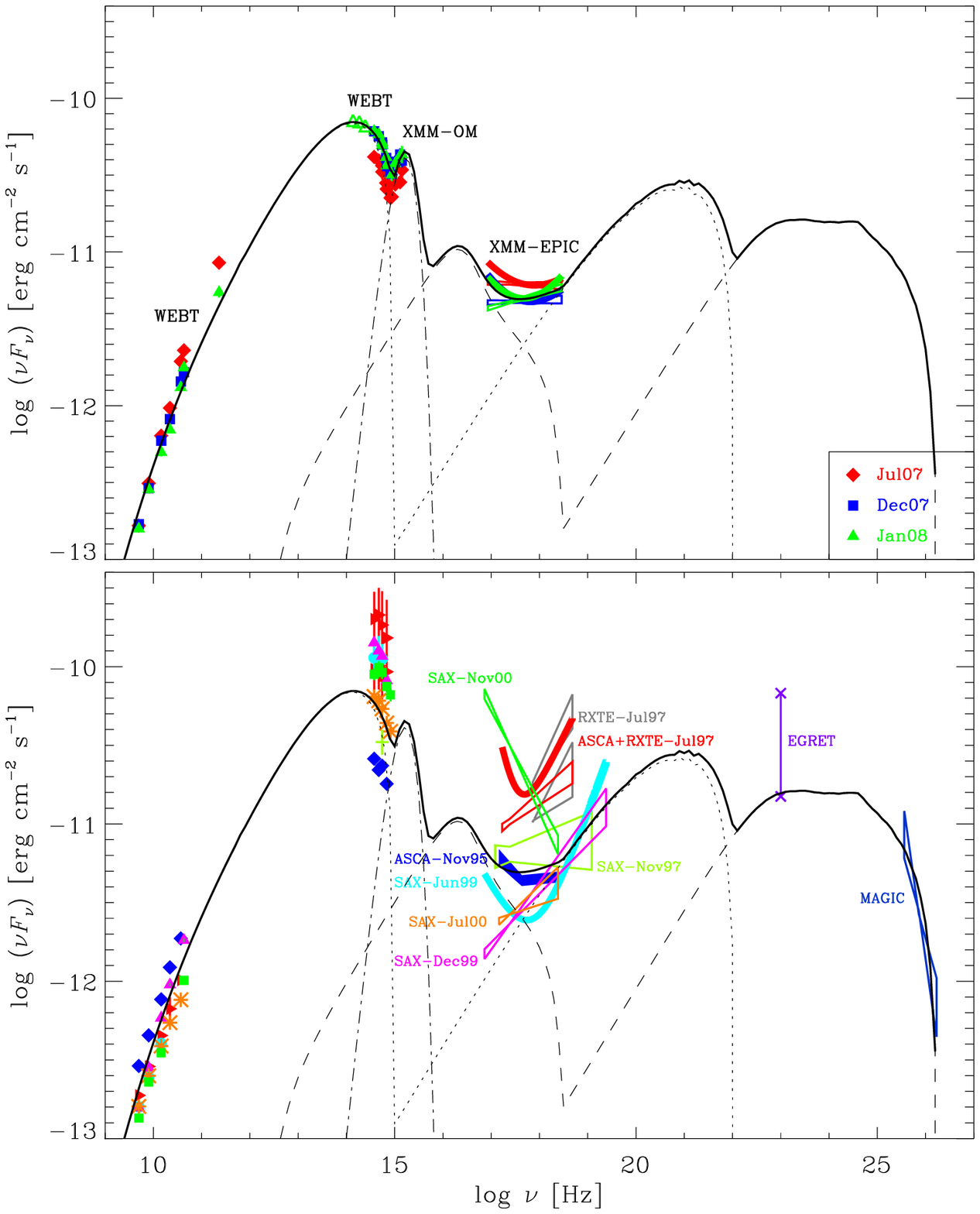}
      \caption{Top panel: broad-band SEDs of BL Lacertae during the three XMM-Newton
 observations of July and December 2007, and January 2008. 
Both the single power law with free absorption and the double power law with atomic plus molecular Galactic absorption fits to the X-ray spectra are shown.
Bottom panel: SEDs corresponding to epochs of previous X-ray satellites observations: 
ASCA in November 1995 \citep[blue,][]{sam99}; 
RXTE in July 1997 \citep[grey,][]{mad99};
ASCA+RXTE in July 1997 \citep[red,][]{tan00}; 
BeppoSAX in November 1997 \citep[light green,][]{pad01};
BeppoSAX in June (cyan) and December (pink) 1999 \citep{rav02};
BeppoSAX in July (orange) and November (green) 2000 \citep{rav03}.
All radio and optical data are from the WEBT archive; the vertical bars on the red optical points indicate the range of variation of the optical fluxes during the ASCA+RXTE observations in July 1997.
The range of flux densities detected by EGRET \citep{har99}, as well as the TeV spectrum observed by MAGIC \citep{alb07}
are also reported.
The solid lines represent the sum of different emission contributions: a low-energy synchrotron+SSC component (dotted lines), a high-energy synchrotron+SSC component (dashed lines), and a thermal component from an accretion disc (dotted-dashed line) with a temperature of $\sim 20000 \rm \, K$ and a luminosity of $6 \times 10^{44} \rm \, erg \, s^{-1}$ (see the text for details).} 
         \label{sed}
   \end{figure*}

The upper panel of Fig.\ \ref{sed} displays the broad-band SEDs corresponding to the XMM-Newton observations analysed in the previous section. We show both the single power law with free absorption and the double power law with atomic plus molecular Galactic absorption fits to the X-ray spectra.
Optical and UV magnitudes were corrected for Galactic extinction by adopting $A_B=1.42$ from \citet{sch98} and
calculating the values at the other wavelengths according to \citet{car89}. De-reddened ground-based optical magnitudes were then converted into fluxes using the zero-mag fluxes given by \citet{bes98}; as for the optical and UV magnitudes from the OM, this conversion was performed using Vega as calibrator.

The optical flux densities were further corrected for the contribution of the host galaxy.
Assuming an $R$-band magnitude of 15.55 after \citet{sca00} and the average colour indices for elliptical galaxies by \citet{man01}, the host galaxy flux densities are: 10.62, 13.97, 11.83, 5.90, 4.23, 2.89, 1.30, and 0.36 mJy in $K, H, J, I, R, V, B$, and $U$ bands, respectively. Using a De Vaucouleurs' profile, we estimated that the contribution to the observed fluxes is $\sim 60$\% of the whole galaxy flux \citep[see also][]{vil02}.
In the UV, the results of spectral evolution modelling of stellar populations by \citet{bru03} allowed us to estimate that the contribution from the host galaxy may be neglected. Indeed, if we consider ages between 4 and 13 Gyr, the galaxy flux density at $\sim 2000$ \AA\ is about 50 to 100 times lower than in the $R$ band, 
i.e.\ the contribution of the galaxy would affect the measured flux in the UV$W2$ band by $\sim 1$--2\%.

Optical data are strictly simultaneous to the XMM-Newton observations. Most radio data are simultaneous too, but in some cases we considered data taken a few days earlier or later. We note the excellent agreement between the $V$, $B$, and $U$ data taken with ground-based telescopes and the corresponding data acquired by the OM.
The main features of the SEDs in the figure are:
\begin{itemize} 
\item the peak of the synchrotron emission lies in the infrared (see below);
\item since a higher radio brightness corresponds to a lower optical  state and viceversa, the synchrotron peak likely shifts toward higher frequencies as the optical flux increases;
\item there is a strong UV excess, since the UV points do not lie on the extrapolation of the optical trend;
\item the X-ray spectrum has either a null slope, or, more likely, it is concave, producing a mild soft-X-ray excess, and suggesting that in this energy region two emission components are intersecting each other; 
\item the optical steepness makes an interpretation of the soft X-ray excess in terms of the tail of the synchrotron component unlikely.
\end{itemize}

\begin{figure}
   \resizebox{\hsize}{!}{\includegraphics{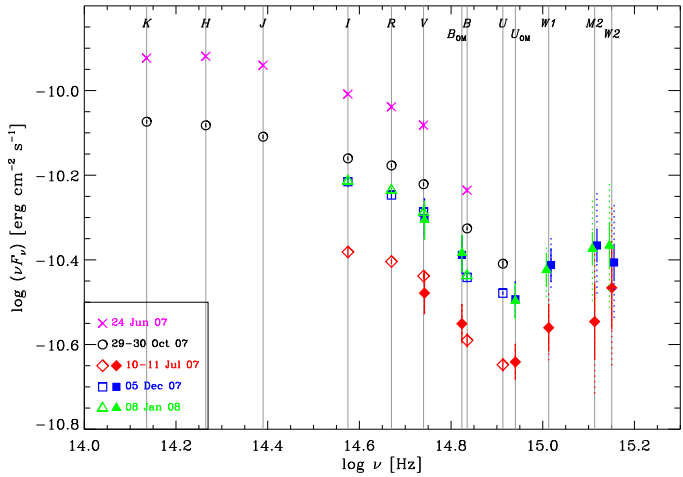}}
      \caption{Details of the near-IR-to-UV SED of BL Lacertae. 
In the three optical--UV SEDs obtained at the epochs 
of the XMM-Newton observations of July 10-11, 2007, December 5, 2007, and January 8, 2008, empty symbols represent ground-based observations, while filled symbols refer to data taken by the OM. The dotted error bars on the UV points indicate the total uncertainty, resulting from considering both the data error (solid error bars) and the sample variance about the mean extinction curve, according to \citet{fit07}. To clearly show the uncertainties holding on the UV data, the UV points corresponding to the December 2007 and January 2008 epochs have been slightly shifted in frequency around the effective value.
For comparison, we show other two SEDs obtained with data acquired by ground-based telescopes on June 24 and October 29-30, 2007, when near-IR data contemporaneous to the optical data were available. The near-IR portion of the SEDs shows that the peak of the synchrotron emission probably lies in this energy range, or close by.} 
         \label{zoom}
   \end{figure}

Details of the near-IR-to-UV SED of BL Lacertae are shown in Fig.\ \ref{zoom}.
Here we can better appreciate the features of the three SEDs obtained at the epochs of the XMM-Newton observations. Moreover, we also report the total uncertainty affecting the UV points, when considering both the error on the data and the sample variance about the Galactic mean extinction curve, following \citet{fit07}.
As stressed by these authors, the sample variance must be taken into account if we want to estimate a realistic error on the de-reddened SEDs. 
In our case, the lower limit to the UV fluxes that we obtain when considering the total uncertainty indicates that the UV excess might be smaller than shown by the points, but it exists, since the UV points cannot be shifted down enough to lie on the extrapolation of the optical, (quasi-power law) synchrotron trend. 

Figure \ref{zoom} also shows two other SEDs built with simultaneous near-IR and optical data.
The near-IR data have been de-reddened and cleaned from the host galaxy contribution similarly to the optical data.
In the optical frequency range, these two SEDs confirm the trend shown by the SEDs obtained at the three XMM-Newton epochs, but they add important information in the near-IR. Indeed, they suggest that the peak of the synchrotron component lies in this energy range, or close by.

\subsection{Comparison with previous observations}

The bottom panel of Fig.\ \ref{sed} shows SEDs corresponding to epochs when various X-ray satellites observed BL Lacertae. The X-ray spectral fits were taken from the literature, while we searched the massive WEBT archive on this source, containing all radio-to-optical data from the previous WEBT campaigns \citep{vil02,vil04a,vil04b,vil09} as well as data from the literature, for contemporaneous low-energy data.

When considering the behaviour of the X-ray spectrum, it is not easy to compare our results to those found by other authors when analysing different X-ray data. The reason is that the shape of the X-ray spectrum strongly depends on the choice of the absorption, i.e.\ on the $N_{\rm H}$ value adopted to perform the spectral fits. Models where the hydrogen column density is left to vary freely yield a variety of $N_{\rm H}$ values (mostly in the range 1.4--$3.5 \times 10^{21} \, \rm cm^{-2}$). These are usually lower than the estimated Galactic total absorption, due to both atomic and molecular hydrogen. The point is that the amount of absorption due to molecular hydrogen is not directly measurable, which introduces a further uncertainty \citep[see Sect.\ 4 and discussion in][]{mad99}. 
In previous analyses of X-ray data, different authors have followed different prescriptions.
An $N_{\rm H}$ value close to the one we assumed was also investigated by \citet{sam99} when analysing ASCA observations in November 1995, and by \citet{pad01} for observations performed by BeppoSAX in November 1997 (and by ROSAT in 1992). In both epochs the X-ray flux was relatively low and \citet{sam99} found that the best fit model was a broken power law, leading to a concave X-ray spectrum.

Observations by RXTE and ASCA during the big outburst of July 1997 were analysed by \citet{mad99} and \citet{tan00}.
The latter authors combined the data of both satellites and distinguished between a low and a high state.
They used a high $N_{\rm H} = 4.6 \times 10^{21} \, \rm cm^{-2}$, and found that the high state was best fitted by a double power law model with an extremely steep spectral index below $\sim 1 \rm \, keV$.
In contrast to this high value, \citet{rav02} and \citet{rav03} preferred an $N_{\rm H} = 2.5 \times 10^{21} \, \rm cm^{-2}$ to analyse BeppoSAX observations in 1999--2000. Nevertheless, the June 1999 data were best fitted by a double power law model, implying a noticeable upward curvature in the soft X-ray domain. In contrast, in November 2000 the whole X-ray spectrum was steep, and its extrapolation to optical frequencies was not intersecting the simultaneous optical spectrum. 
This offset was questioned by \citet{boe03}, who warned about the possibility that it was produced by flux averaging in a period of noticeable variability, as shown by the intensive monitoring performed by the WEBT observers during the 2000--2001 observing season \citep{vil02}.
However, to remove the offset we should have missed one or more big flares, such as to increase the mean optical level by more than 1 mag, which seems unlikely.
Hence, if in previous works a soft spectral component at low X-ray energies had always been interpreted as the tail of the synchrotron emission, this was no longer a plausible explanation for the November 2000 SED, as it is not a plausible explanation for the XMM-Newton SEDs presented above.
A number of hypotheses were suggested by \citet{rav03} to justify the optical--X-ray offset: a sudden increase of the dust-to-gas ratio toward BL Lacertae; the detection of a bulk Compton emission; the interplay of two different synchrotron components; Klein-Nishina effect on the synchrotron spectrum.

\section{The helical jet model}

The availability of broad-band SEDs built with simultaneous data, including UV information, at the epochs of the XMM-Newton observations of 2007--2008 is a formidable tool to investigate the nature of BL Lac multiwavelength emission. The picture that we described in the previous section suggests that the SED behaviour cannot be explained in terms of one synchrotron plus its SSC emission components.

Thermal emission from an accretion disc could account for the UV excess.
Indeed, the optical spectrum of this source occasionally shows broad emission lines that are produced in the broad line region, which is most likely photoionised by the radiation coming from the accretion disc \citep{ver95,cor96,cor00}. 
%The optical spectra that we acquired with the TNG clearly show the broad H$\alpha$ emission line (Raiteri et al., in preparation), implying that in the 2007--2008 observing season the accretion disc was luminous.
Moreover, a big blue bump produced in the SED by thermal emission from an accretion disc has already been found for other quasar-type blazars, like 3C 273 \citep{smi93,von97,gra04,tur06}, 3C 279 \citep{pia99}, 3C 345 \citep{bre86}, and 3C 454.3 \citep{rai07b,rai08c}.

However, a thermal emission component would hardly be able to explain the extremely variable X-ray spectrum shown in the bottom panel of Fig.\ \ref{sed}. 
Indeed, at 0.3 keV BeppoSAX observed a flux variation of a factor $\sim 50$ in about 1 year.
We notice that the behaviour of BL Lacertae in the UV--X-ray energy range is similar to that found by \citet{rai05,rai06a,rai06b} when analysing the SEDs of another BL Lac object, AO 0235+164, which also occasionally shows broad emission lines \citep[see e.g.][and references therein]{rai07a}. In that case, the alternative hypothesis of a second, higher-frequency synchrotron component, likely coming from an inner region of the jet, was left open.

We investigated this problem by means of the helical jet model by \citet[][see also \citealt{rai99}, \citealt{rai03p}, and \citealt{ost04}]{vil99}.
We tried to reproduce the ``XMM-Newton" broad-band SEDs of December 2007 and January 2008, which are very similar. To better trace the SED shape, we also added synthetic near-IR data, derived by shifting the October 2007 SED of Fig.\ \ref{zoom} to match the above ``XMM-Newton" SEDs. This implied a shift of $\log (\nu F_\nu) = -0.07$.

The helical jet model presented in \citet{vil99} foresees that orbital motion in a
binary black hole system, coupled with the interaction of the plasma jet with the surrounding medium, 
twists the emitting jet in a rotating helical structure. 
We here recall the main features of the model.
The axis of the helical-shaped jet is assumed to lie
along the $z$-axis of a 3-D reference frame. The pitch angle is
$\zeta$ and $\psi$ is the angle defined by the helix axis with the
line of sight. 
The non-dimensional length of the helical path can be expressed in terms
of the $z$ coordinate along the helix axis:
\begin{equation} l(z)=\frac{z}{\cos \zeta}\,,\quad 0\le z \le 1\,,  \end{equation}
which corresponds to an azimuthal angle 
$\varphi(z)=az$, where the angle $a$ is a constant.
The jet viewing angle varies along the helical path as
\begin{equation}			\label{costeta}
\cos\theta(z)=\cos\psi\cos\zeta+\sin\psi\sin\zeta\cos(\phi- a z)\,,
\end{equation}
where $\phi$ is the azimuthal difference between the line of sight and the
initial direction of the helical path.

The jet is inhomogeneous: it emits radiation at progressively
increasing wavelengths by proceeding from its apex outwards.
Each slice of the jet can radiate, in the plasma rest
reference frame, synchrotron photons from a minimum
frequency $\nu'_{\mathrm{s,min}}$ to a maximum one $\nu'_{\mathrm{s,max}}$. 
Both these frequency limits decrease for increasing distance from the jet apex
following a power law:
\begin{equation}		 \label{nusyn}
\nu'_{{\rm s},i}(l)=\nu'_{\mathrm{s}}(0)
\left(1+\frac{l}{l_i}\right)^{-c_i} \, , \quad c_i>0\,, \end{equation}
where $l_i$ are length scales, and $i=\rm min,max$.
The high-energy emission producing the second bump in the blazars SED is
assumed to be the result of inverse Compton scattering of the synchrotron photons
by the same relativistic electrons emitting them (SSC model). 
Consequently, each portion of the jet emitting synchrotron radiation between 
$\nu'_{\mathrm{s,min}}(l)$ and $\nu'_{\mathrm{s,max}}(l)$ will also produce inverse
Compton radiation between $\nu'_{\mathrm{c,min}}(l)$ and $\nu'_{\mathrm{c,max}}(l)$,
with $\nu'_{{\rm c},i}(l)= \frac{4}{3}\gamma_i^{2}(l) \nu'_{{\rm s},i}(l)$.
The electron Lorentz factor ranges from $\gamma_{\rm min}=1$ to $\gamma_{\rm max}(l)$, 
which has a similar power law dependence as in Eq.\ \ref{nusyn}, 
with power $c_\gamma$ and length scale $l_\gamma$.
%$c_\gamma=0.5 \, c_{\rm max}$ and $l_\gamma= l_{\rm max}$.
%\footnote{In the jet model by \citet{ghi89} these relations would imply that the magnetic field strength is constant.}
As photon energies increase, the classical Thomson scattering cross
section is replaced by the Klein-Nishina one, which takes into
account quantum effects. Its consequence is to reduce the cross section
from its classical value, so that Compton 
scattering becomes less efficient at high energies. 
We approximated this effect by requiring that  
$\nu'_\mathrm{c,max}(l)$ be averaged with
$\nu'^{\rm KN}_\mathrm{c,max}(l)=\frac{m_{\mathrm{e}} c^{2}}{h} \, \gamma_\mathrm{max}(l)$ 
when $\gamma_\mathrm{max}(l)\nu'_\mathrm{s,max}(l) >
\frac{3}{4} \frac{m_{\mathrm{e}} c^{2}}{h}$.

We assume a power law dependence of the observed flux density on the frequency and
a cubic dependence on the Doppler beaming factor $\delta$:
$F_\nu(\nu)\propto \delta^3 \nu^{-\alpha_0}$, 
where $\alpha_0$ is the power law index of the local synchrotron spectrum,
$\delta =[\Gamma(1-\beta\cos\theta)]^{-1}$,
$\beta$ is the bulk velocity of the emitting plasma
in units of the speed of light, $\Gamma=(1-\beta^2)^{-1/2}$ the corresponding
bulk Lorentz factor, and $\theta$ is the viewing angle of Eq.\ \ref{costeta}.
Since the viewing angle varies along the helical path, also the beaming factor does.
Hence, the flux at $\nu$ peaks when the part of the jet mostly
contributing to it has minimum $\theta$.

The emissivity decreases along the jet: both the synchrotron and inverse
Compton flux densities are allowed to drop when moving from the jet apex
outwards. For a jet slice of thickness d$l$:
\begin{equation}
{\rm d}F_{\nu,{\rm s}}(\nu) \propto \, \delta^3(l) \, \nu^{-\alpha_0} \,
\left(1+\frac{l}{l_{\rm s}}\right)^{-c_{\rm s}} {\rm d}l \, , \quad c_{\rm s}>0 \, ,
\end{equation}
\begin{equation}
{\rm d}F_{\nu,{\rm c}}(\nu) \propto \, \delta^3(l) \, \nu^{-\alpha_0} \,
\left(1+\frac{l}{l_{\rm c}}\right)^{-c_{\rm c}}
\ln \left[\frac{\nu'_{\rm s,max}(l)}{\nu'_{\rm s,min}(l)}\right]  
{\rm d}l \, , \quad c_{\rm c}>0 \,  
\end{equation}

For both the synchrotron and inverse Compton components, the observed flux densities at frequency $\nu$ coming from the whole jet are obtained by integrating over all the jet portions
$\Delta z_i (\nu)$ contributing to that observed frequency, i.e.\ for which  
$\delta(z)\nu'_\mathrm{min}(z)\leq\nu\leq\delta(z)\nu'_\mathrm{max}(z)$.
The total observed flux density at frequency $\nu$ is finally obtained by 
summing the synchrotron and inverse Compton contributions.
Notice that the intrinsic jet emission does not vary with time, but the observed one may change as the orientation changes.

The fit to the ``XMM-Newton" broad-band SED of December 2007 - January 2008 in Fig.\ \ref{sed} (solid line) is obtained by considering two synchrotron emission components from different regions of a helical jet, with their corresponding SSC, plus a thermal component, modelled as a black body. 
The main parameters of the model are reported in table \ref{para}.
\begin{table}
\caption{Main parameters of the helical model for the fit to the ``XMM-Newton" broad-band SED of December 2007 - January 2008. The differences between the low- and high-energy synchrotron+SSC components are easily seen. }
\label{para}
\centering
\begin{tabular}{l r r  }
\hline\hline
Parameter & Low & High \\
\hline
$\zeta$    & $30 \degr$ & $30 \degr$\\
$\psi$     & $25 \degr$ & $25 \degr$ \\
$a$        & $110 \degr$ & $110 \degr$\\
$\phi$     & $-8 \degr$ & $20 \degr$ \\
$\log \nu'_{\rm s}(0)$ & 14.0 & 17.8 \\
$c_{\rm min,max}$ & 2.5 & 2.5\\
$\log l_{\rm min}$ & $-3.2$ & $-3.2$\\
$\log l_{\rm max}$ & $-1.6$ & $-1.6$\\
$\log \gamma_{\rm max}(0)$ & 3.5 & 4.4 \\
$c_\gamma$ & 1.25 & 1.25 \\
$\log l_\gamma$ & $-1.6$ & $-1.6$\\
$\alpha_0$ & 0.5 & 0.5 \\
$\Gamma$   & 10 & 10\\
$c_{\rm s}$ & 1 & 1\\
$\log l_{\rm s}$ & $-1$ & $-1$\\
$c_{\rm c}$ & 1 & 1\\
$\log l_{\rm c}$ & $-1$ & $-1$\\
\hline
\end{tabular}
\end{table}

%The helix pitch angle is $30 \degr$, and the helix axis is at $25 \degr$ from the line of sight. 
%The index of the power law dependence of the flux on frequency is $\alpha_0=0.5$. The bulk Lorentz factor is 10.
The lower-energy synchrotron+SSC emission (dotted line) comes from a helical portion that is initially fairly aligned with the line of sight ($\phi=-8 \degr$, $\theta (0) \approx 6.2 \degr$), while the higher-energy synchrotron+SSC emission (dashed line) is produced by another helical region that is initially less aligned with the line of sight  ($\phi=20 \degr$, $\theta (0) \approx 10.4 \degr$).

A thermal component that fits the UV excess must have a black body temperature $\ga 20000 \rm \, K$ and a luminosity $\ga 6 \times 10^{44} \, \rm erg \, s^{-1}$. This lower limit to the temperature (and consequently to the luminosity) is constrained by the break of the SED in the optical--UV transition, but much hotter and hence more luminous discs are possible.
For comparison, the thermal disc fitted by \citet{pia99} to the UV data of 3C 279
has a temperature of 20000 K and a luminosity of $2 \times 10^{45} \, \rm erg \, s^{-1}$.
%The luminosity would be smaller in case the Galactic extinction has been overestimated.

The two synchrotron+SSC components can change a lot for variations e.g.\ of the angle $\phi$, which happens if the helix rotates, allowing us to explain the noticeable spectral variability of the source also in the absence of intrinsic, energetic processes.
With reference to the SEDs shown in the bottom panel of Fig.\ \ref{sed}, the high-energy synchrotron+SSC emission was giving an exceptional contribution in November 2000, while it was very faint in December 1999.
A detailed investigation of the model parameter space to fit the SED shape of BL Lacertae at different epochs goes beyond the scope of this paper. 
We notice however that our model fit produces a GeV spectrum with photon index $\Gamma \approx 2$ and can fairly reproduce the TeV spectrum observed by the MAGIC telescope in 2005 \citep{alb07}.

\section{Discussion and conclusions}

The WEBT campaign on BL Lacertae in the 2007--2008 observing season involved 37 optical-to-radio telescopes. They observed the source in a relatively faint state.
Nevertheless, some fast variability episodes were detected in the optical bands, superposed to a long-term flux increasing trend. 
During the campaign, three observations by the XMM-Newton satellite added information on the UV and X-ray states of the source.

The broad-band SEDs built with simultaneous data taken at the epochs of the XMM-Newton observations show a clear UV excess. 
The high UV fluxes are explained if we assume a contribution by thermal radiation from the accretion disc. 
%This hypothesis is confirmed by the analysis of the optical spectra obtained in the same period with the 3.6 m Telescopio Nazionale Galileo (TNG, Raiteri et al, in preparation), which show both a broad H$\alpha$ and several narrow emission lines, as already found in 1995--1997 by \citet{ver95} and \citet{cor96,cor00}.

On the other side, the corresponding X-ray spectra indicate a possible soft excess. When comparing our X-ray data with previous data from other satellites, the X-ray spectrum appears to vary dramatically, so the soft excess cannot be ascribed to the accretion disc and/or hot corona surrounding it. 
A much more variable emission contribution is required.

We found that the broad-band SEDs of BL Lacertae can be explained in terms of two synchrotron emission components with their corresponding SSC radiation, plus a thermal component representing the contribution of the accretion disc. 
When fitting these two non-thermal components by means of the helical jet model of \citet{vil99}, and the thermal one with a black body law, we find that the accretion disc has a temperature $\ga 20000 \, \rm K$ and a luminosity $\ga 6 \times 10^{44} \, \rm erg \, s^{-1}$. Taking into account that $L=\eta \dot{M} c^2$, with $\eta \simeq 0.06$ in the case of Schwarzschild's metric \citep{sha73}, we can derive a lower limit to the accretion rate: $\dot{M} \ga 0.2 \, M_{\sun} \,  \rm yr^{-1}$.
And if we assume that the luminosity equals the Eddington's critical luminosity, we can also infer a lower limit to the black hole mass: $M_{\rm BH} \ga 6 \times 10^6 M_{\sun}$. This value scales as $L_{\rm Edd}/L$.

A further emission component coming from inverse-Compton scattering on external photons from the disc and/or the broad line region might be present, but it is not needed to fit the observations, so we think that its possible contribution would be a minor one. Its modelling would introduce additional parameters that we could not reliably constrain. 

Being aware that also other interpretations might be able to account for the observations reported in this paper, we notice that a helical jet model is motivated by some observing evidence. Indeed, VLBA/VLBI studies of the jet structure in AGNs have revealed bent jet morphologies that are suggestive of streaming motions along a helical path \citep[see e.g.][]{lis01} or that the magnetic field may present a helical geometry \citep{gab04}; this is also true for BL Lacertae \citep[][see also \citealt{sti03}]{tat98,den00,mar08,osu09}.
Moreover, a rotating helical path in a curved jet was invoked by \citet{vil09} to explain the optical and radio behaviour of BL Lacertae in the last forty years, in particular the alternation of enhanced and suppressed optical activity, accompanied by hard and soft radio events, respectively.
If the jet has a helical structure, different portions of the jet may be well aligned with the line of sight, with consequent Doppler beaming of the emitted radiation.
The two emission components proposed in our modelling could correspond to two of these regions, the higher-energy one being located closer to the jet apex.
An analogous picture was suggested by \citet{vil99} for Mkn 501, whose radio-to-X-ray multiepoch SED was explained in terms of two jet regions with different curvature.
Alternatively, we can imagine that the two contributions come from two interweaved helical filaments, but we regard this hypothesis as less likely.

Finally, we mention that also in the case of Mkn 421, a possible explanation for its X-ray and optical flux behaviour during the June 2008 flare implies the existence of two different synchrotron emitting regions in the jet \citep{don09a}.

Further multiwavelength observations, including GeV data from the Fermi satellite, and TeV data from ground-based Cherenkov telescopes, will help verify our interpretation.

\begin{acknowledgements}
This work is partly based on observations made with the Nordic Optical Telescope, operated
on the island of La Palma jointly by Denmark, Finland, Iceland,
Norway, and Sweden, in the Spanish Observatorio del Roque de los
Muchachos of the Instituto de Astrofisica de Canarias, and on observations collected at the German-Spanish Calar Alto Observatory, jointly operated by the MPIA and the IAA-CSIC.
AZT-24 observations are made within an agreement between  Pulkovo, Rome and Teramo observatories.
The Submillimeter Array is a joint project between the Smithsonian Astrophysical Observatory and the Academia Sinica Institute of Astronomy and Astrophysics and is funded by the Smithsonian Institution and the Academia Sinica.
This research has made use of data from the University of Michigan Radio Astronomy Observatory,
which is supported by the National Science Foundation and by funds from the University of Michigan.
This work is partly based on observation from Medicina and Noto telescopes operated by INAF - Istituto di Radioastronomia.
The Torino team acknowledges financial support by the Italian Space Agency through contract ASI-INAF I/088/06/0 for the Study of High-Energy Astrophysics.
Acquisition of the MAPCAT data at the Calar Alto Observatory is supported in part by the Spanish ``Ministerio de Ciencia e Innovaci\'on" through grant AYA2007-67626-C03-03.
The Mets\"ahovi team acknowledges the support from the Academy of Finland.
This research was partially supported by Scientific Research Fund of the
Bulgarian Ministry of Education and Sciences (BIn - 13/09).
St.Petersburg University team acknowledges support from RFBR grant 09-02-00092.
Observations at the Abastumani 70-cm meniscus were partially supported by the
Georgian National Science Foundation grant GNSF/ST-08/4-404.
This research has made use of NASA's Astrophysics Data System.
\end{acknowledgements}


\begin{thebibliography}{72}
\expandafter\ifx\csname natexlab\endcsname\relax\def\natexlab#1{#1}\fi

\bibitem[{{Albert} {et~al.}(2007){Albert}, {Aliu}, {Anderhub}, {Antoranz},
  {Armada}, {Baixeras}, {Barrio}, {Bartko}, {Bastieri}, {Becker}, {Bednarek},
  {Berger}, {Bigongiari}, {Biland}, {Bock}, {Bordas}, {Bosch-Ramon}, {Bretz},
  {Britvitch}, {Camara}, {Carmona}, {Chilingarian}, {Coarasa}, {Commichau},
  {Contreras}, {Cortina}, {Costado}, {Curtef}, {Danielyan}, {Dazzi}, {De
  Angelis}, {Delgado}, {de los Reyes}, {De Lotto}, {Domingo-Santamar{\'{\i}}a},
  {Dorner}, {Doro}, {Errando}, {Fagiolini}, {Ferenc}, {Fern{\'a}ndez}, {Firpo},
  {Flix}, {Fonseca}, {Font}, {Fuchs}, {Galante}, {Garc{\'{\i}}a-L{\'o}pez},
  {Garczarczyk}, {Gaug}, {Giller}, {Goebel}, {Hakobyan}, {Hayashida},
  {Hengstebeck}, {Herrero}, {H{\"o}hne}, {Hose}, {Hsu}, {Jacon}, {Jogler},
  {Kosyra}, {Kranich}, {Kritzer}, {Laille}, {Lindfors}, {Lombardi}, {Longo},
  {L{\'o}pez}, {L{\'o}pez}, {Lorenz}, {Majumdar}, {Maneva}, {Mannheim},
  {Mansutti}, {Mariotti}, {Mart{\'{\i}}nez}, {Mazin}, {Merck}, {Meucci},
  {Meyer}, {Miranda}, {Mirzoyan}, {Mizobuchi}, {Moralejo}, {Nilsson},
  {Ninkovic}, {O{\~n}a-Wilhelmi}, {Otte}, {Oya}, {Paneque}, {Panniello},
  {Paoletti}, {Paredes}, {Pasanen}, {Pascoli}, {Pauss}, {Pegna}, {Persic},
  {Peruzzo}, {Piccioli}, {Poller}, {Prandini}, {Puchades}, {Raymers}, {Rhode},
  {Rib{\'o}}, {Rico}, {Rissi}, {Robert}, {R{\"u}gamer}, {Saggion},
  {S{\'a}nchez}, {Sartori}, {Scalzotto}, {Scapin}, {Schmitt}, {Schweizer},
  {Shayduk}, {Shinozaki}, {Shore}, {Sidro}, {Sillanp{\"a}{\"a}}, {Sobczynska},
  {Stamerra}, {Stark}, {Takalo}, {Temnikov}, {Tescaro}, {Teshima}, {Tonello},
  {Torres}, {Turini}, {Vankov}, {Vitale}, {Wagner}, {Wibig}, {Wittek},
  {Zandanel}, {Zanin}, \& {Zapatero}}]{alb07}
{Albert}, J., {Aliu}, E., {Anderhub}, H., {et~al.} 2007, \apjl, 666, L17

\bibitem[{{Bach} {et~al.}(2006){Bach}, {Villata}, {Raiteri}, {Agudo}, {Aller},
  {Aller}, {Denn}, {G{\'o}mez}, {Jorstad}, {Marscher}, {Mutel}, \&
  {Ter{\"a}sranta}}]{bac06}
{Bach}, U., {Villata}, M., {Raiteri}, C.~M., {et~al.} 2006, \aap, 456, 105

\bibitem[{{Bania} {et~al.}(1991){Bania}, {Marscher}, \& {Barvainis}}]{ban91}
{Bania}, T.~M., {Marscher}, A.~P., \& {Barvainis}, R. 1991, \aj, 101, 2147

\bibitem[{{Bertaud} {et~al.}(1969){Bertaud}, {Dumortier}, {Veron}, {Wlerick},
  {Adam}, {Bigay}, {Garnier}, \& {Duruy}}]{ber69}
{Bertaud}, C., {Dumortier}, B., {Veron}, P., {et~al.} 1969, \aap, 3, 436

\bibitem[{{Bessell} {et~al.}(1998){Bessell}, {Castelli}, \& {Plez}}]{bes98}
{Bessell}, M.~S., {Castelli}, F., \& {Plez}, B. 1998, \aap, 333, 231

\bibitem[{{Bloom} {et~al.}(1997){Bloom}, {Bertsch}, {Hartman}, {Sreekumar},
  {Thompson}, {Balonek}, {Beckerman}, {Davis}, {Whitman}, {Miller}, {Nair},
  {Roberts}, {Tosti}, {Massaro}, {Nesci}, {Maesano}, {Montagni}, {Jang},
  {Bock}, {Dietrich}, {Herter}, {Otterbein}, {Pfeiffer}, {Seitz}, \&
  {Wagner}}]{blo97}
{Bloom}, S.~D., {Bertsch}, D.~L., {Hartman}, R.~C., {et~al.} 1997, \apjl, 490,
  L145

\bibitem[{{B{\"o}ttcher} \& {Bloom}(2000)}]{boe00}
{B{\"o}ttcher}, M. \& {Bloom}, S.~D. 2000, \aj, 119, 469

\bibitem[{{B{\"o}ttcher} {et~al.}(2003){B{\"o}ttcher}, {Marscher}, {Ravasio},
  {Villata}, {Raiteri}, {Aller}, {Aller}, {Ter{\"a}sranta}, {Mang},
  {Tagliaferri}, {Aharonian}, {Krawczynski}, {Kurtanidze}, {Nikolashvili},
  {Ibrahimov}, {Papadakis}, {Tsinganos}, {Sadakane}, {Okada}, {Takalo},
  {Sillanp{\"a}{\"a}}, {Tosti}, {Ciprini}, {Frasca}, {Marilli}, {Robb},
  {Noble}, {Jorstad}, {Hagen-Thorn}, {Larionov}, {Nesci}, {Maesano},
  {Schwartz}, {Basler}, {Gorham}, {Iwamatsu}, {Kato}, {Pullen},
  {Ben{\'{\i}}tez}, {de Diego}, {Moilanen}, {Oksanen}, {Rodriguez}, {Sadun},
  {Kelly}, {Carini}, {Miller}, {Catalano}, {Dultzin-Hacyan}, {Fan},
  {Ghisellini}, {Ishioka}, {Karttunen}, {Kein{\"a}nen}, {Kudryavtseva},
  {Lainela}, {Lanteri}, {Larionova}, {Matsumoto}, {Mattox}, {McHardy},
  {Montagni}, {Nucciarelli}, {Ostorero}, {Papamastorakis}, {Pasanen},
  {Sobrito}, \& {Uemura}}]{boe03}
{B{\"o}ttcher}, M., {Marscher}, A.~P., {Ravasio}, M., {et~al.} 2003, \apj, 596,
  847

\bibitem[{{Bregman} {et~al.}(1986){Bregman}, {Glassgold}, {Huggins},
  {Neugebauer}, {Soifer}, {Matthews}, {Elias}, {Webb}, {Pollock}, {Pica},
  {Leacock}, {Smith}, {Aller}, {Aller}, {Hodge}, {Dent}, {Balonek},
  {Barvainis}, {Roellig}, {Wisniewski}, {Rieke}, {Lebofsky}, {Wills}, {Wills},
  {Ku}, {Bregman}, {Witteborn}, {Lester}, {Impey}, \& {Hackwell}}]{bre86}
{Bregman}, J.~N., {Glassgold}, A.~E., {Huggins}, P.~J., {et~al.} 1986, \apj,
  301, 708

\bibitem[{{Bruzual} \& {Charlot}(2003)}]{bru03}
{Bruzual}, G. \& {Charlot}, S. 2003, \mnras, 344, 1000

\bibitem[{{Cardelli} {et~al.}(1989){Cardelli}, {Clayton}, \& {Mathis}}]{car89}
{Cardelli}, J.~A., {Clayton}, G.~C., \& {Mathis}, J.~S. 1989, \apj, 345, 245

\bibitem[{{Chen} {et~al.}(2008){Chen}, {D'Ammando}, {Villata}, {Raiteri},
  {Tavani}, {Vittorini}, {Bulgarelli}, {Donnarumma}, {Ferrari}, {Giuliani},
  {Longo}, {Pacciani}, {Pucella}, {Vercellone}, {Argan}, {Barbiellini},
  {Boffelli}, {Caraveo}, {Carosati}, {Cattaneo}, {Cocco}, {Costa}, {Del Monte},
  {de Paris}, {di Cocco}, {Evangelista}, {Feroci}, {Fiorini}, {Froysland},
  {Frutti}, {Fuschino}, {Galli}, {Gianotti}, {Kurtanidze}, {Labanti},
  {Lapshov}, {Larionov}, {Lazzarotto}, {Lipari}, {Marisaldi}, {Mastropietro},
  {Mereghetti}, {Morelli}, {Morselli}, {Pasanen}, {Pellizzoni}, {Perotti},
  {Picozza}, {Porrovecchio}, {Prest}, {Rapisarda}, {Rappoldi}, {Rubini},
  {Soffitta}, {Trifoglio}, {Trois}, {Vallazza}, {Zambra}, {Zanello}, {Cutini},
  {Gasparrini}, {Pittori}, {Santolamazza}, {Verrecchia}, {Giommi}, {Antonelli},
  {Colafrancesco}, \& {Salotti}}]{che08}
{Chen}, A.~W., {D'Ammando}, F., {Villata}, M., {et~al.} 2008, \aap, 489, L37

\bibitem[{{Corbett} {et~al.}(2000){Corbett}, {Robinson}, {Axon}, \&
  {Hough}}]{cor00}
{Corbett}, E.~A., {Robinson}, A., {Axon}, D.~J., \& {Hough}, J.~H. 2000,
  \mnras, 311, 485

\bibitem[{{Corbett} {et~al.}(1996){Corbett}, {Robinson}, {Axon}, {Hough},
  {Jeffries}, {Thurston}, \& {Young}}]{cor96}
{Corbett}, E.~A., {Robinson}, A., {Axon}, D.~J., {et~al.} 1996, \mnras, 281,
  737

\bibitem[{{D'Ammando} {et~al.}(2009){D'Ammando}, {Pucella}, {Raiteri},
  {Villata}, {Vittorini}, {Vercellone}, {Donnarumma}, \& {et al.}}]{dam09}
{D'Ammando}, F., {Pucella}, G., {Raiteri}, C.~M., {et~al.} 2009, \aap,
  submitted

\bibitem[{{Denn} {et~al.}(2000){Denn}, {Mutel}, \& {Marscher}}]{den00}
{Denn}, G.~R., {Mutel}, R.~L., \& {Marscher}, A.~P. 2000, \apjs, 129, 61

\bibitem[{{Dermer} {et~al.}(1992){Dermer}, {Schlickeiser}, \&
  {Mastichiadis}}]{der92}
{Dermer}, C.~D., {Schlickeiser}, R., \& {Mastichiadis}, A. 1992, \aap, 256, L27

\bibitem[{{Donnarumma} {et~al.}(2009{\natexlab{a}}){Donnarumma}, {Pucella},
  {Vittorini}, {D'Ammando}, {Vercellone}, {Raiteri}, {Villata}, \& {et
  al.}}]{don09b}
{Donnarumma}, I., {Pucella}, G., {Vittorini}, V., {et~al.} 2009{\natexlab{a}},
  \apj, submitted

\bibitem[{{Donnarumma} {et~al.}(2009{\natexlab{b}}){Donnarumma}, {Vittorini},
  {Vercellone}, {Monte}, {Feroci}, {D'Ammando}, {Pacciani}, {Chen}, {Tavani},
  {Bulgarelli}, {Giuliani}, {Longo}, {Pucella}, {Argan}, {Barbiellini},
  {Boffelli}, {Caraveo}, {Cattaneo}, {Cocco}, {Costa}, {DeParis}, {Cocco},
  {Evangelista}, {Fiorini}, {Froysland}, {Frutti}, {Fuschino}, {Galli},
  {Gianotti}, {Labanti}, {Lapshov}, {Lazzarotto}, {Lipari}, {Marisaldi},
  {Mastropietro}, {Mereghetti}, {Morelli}, {Morselli}, {Pellizzoni}, {Perotti},
  {Picozza}, {Porrovecchio}, {Prest}, {Rapisarda}, {Rappoldi}, {Rubini},
  {Soffitta}, {Trifoglio}, {Trois}, {Vallazza}, {Zambra}, {Zanello}, {Pittori},
  {Santolamazza}, {Verrecchia}, {Giommi}, {Colafrancesco}, {Salotti},
  {Villata}, {Raiteri}, {Chen}, {Efimova}, {Jordan}, {Konstantinova},
  {Koptelova}, {Kurtanidze}, {Larionov}, {Ros}, {Sadun}, \& {et al.}}]{don09a}
{Donnarumma}, I., {Vittorini}, V., {Vercellone}, S., {et~al.}
  2009{\natexlab{b}}, \apjl, 691, L13

\bibitem[{{Fiorucci} \& {Tosti}(1996)}]{fio96}
{Fiorucci}, M. \& {Tosti}, G. 1996, \aaps, 116, 403

\bibitem[{{Fitzpatrick} \& {Massa}(2007)}]{fit07}
{Fitzpatrick}, E.~L. \& {Massa}, D. 2007, \apj, 663, 320

\bibitem[{{Gabuzda} {et~al.}(2004){Gabuzda}, {Murray}, \& {Cronin}}]{gab04}
{Gabuzda}, D.~C., {Murray}, {\'E}., \& {Cronin}, P. 2004, \mnras, 351, L89

\bibitem[{{Ghisellini} \& {Tavecchio}(2009)}]{ghi09}
{Ghisellini}, G. \& {Tavecchio}, F. 2009, \mnras, 397, 985

\bibitem[{{Grandi} \& {Palumbo}(2004)}]{gra04}
{Grandi}, P. \& {Palumbo}, G.~G.~C. 2004, Science, 306, 998

\bibitem[{{Hartman} {et~al.}(1999){Hartman}, {Bertsch}, {Bloom}, {Chen},
  {Deines-Jones}, {Esposito}, {Fichtel}, {Friedlander}, {Hunter}, {McDonald},
  {Sreekumar}, {Thompson}, {Jones}, {Lin}, {Michelson}, {Nolan}, {Tompkins},
  {Kanbach}, {Mayer-Hasselwander}, {M{\"u}cke}, {Pohl}, {Reimer}, {Kniffen},
  {Schneid}, {von Montigny}, {Mukherjee}, \& {Dingus}}]{har99}
{Hartman}, R.~C., {Bertsch}, D.~L., {Bloom}, S.~D., {et~al.} 1999, \apjs, 123,
  79

\bibitem[{{Kalberla} {et~al.}(2005){Kalberla}, {Burton}, {Hartmann}, {Arnal},
  {Bajaja}, {Morras}, \& {P{\"o}ppel}}]{kal05}
{Kalberla}, P.~M.~W., {Burton}, W.~B., {Hartmann}, D., {et~al.} 2005, \aap,
  440, 775

\bibitem[{{Lister}(2001)}]{lis01}
{Lister}, M.~L. 2001, \apj, 562, 208

\bibitem[{{Liszt}(2007)}]{lis07}
{Liszt}, H.~S. 2007, \aap, 476, 291

\bibitem[{{Liszt} \& {Lucas}(1998)}]{lis98}
{Liszt}, H.~S. \& {Lucas}, R. 1998, \aap, 339, 561

\bibitem[{{Lucas} \& {Liszt}(1993)}]{luc93}
{Lucas}, R. \& {Liszt}, H.~S. 1993, \aap, 276, L33+

\bibitem[{{Madejski} {et~al.}(1999){Madejski}, {Sikora}, {Jaffe},
  {B{\l}a{\.z}ejowski}, {Jahoda}, \& {Moderski}}]{mad99}
{Madejski}, G.~M., {Sikora}, M., {Jaffe}, T., {et~al.} 1999, \apj, 521, 145

\bibitem[{{Mannucci} {et~al.}(2001){Mannucci}, {Basile}, {Poggianti},
  {Cimatti}, {Daddi}, {Pozzetti}, \& {Vanzi}}]{man01}
{Mannucci}, F., {Basile}, F., {Poggianti}, B.~M., {et~al.} 2001, \mnras, 326,
  745

\bibitem[{{Marscher} {et~al.}(2008){Marscher}, {Jorstad}, {D'Arcangelo},
  {Smith}, {Williams}, {Larionov}, {Oh}, {Olmstead}, {Aller}, {Aller},
  {McHardy}, {L{\"a}hteenm{\"a}ki}, {Tornikoski}, {Valtaoja}, {Hagen-Thorn},
  {Kopatskaya}, {Gear}, {Tosti}, {Kurtanidze}, {Nikolashvili}, {Sigua},
  {Miller}, \& {Ryle}}]{mar08}
{Marscher}, A.~P., {Jorstad}, S.~G., {D'Arcangelo}, F.~D., {et~al.} 2008, \nat,
  452, 966

\bibitem[{{Mason} {et~al.}(2001){Mason}, {Breeveld}, {Much}, {Carter},
  {Cordova}, {Cropper}, {Fordham}, {Huckle}, {Ho}, {Kawakami}, {Kennea},
  {Kennedy}, {Mittaz}, {Pandel}, {Priedhorsky}, {Sasseen}, {Shirey}, {Smith},
  \& {Vreux}}]{mas01}
{Mason}, K.~O., {Breeveld}, A., {Much}, R., {et~al.} 2001, \aap, 365, L36

\bibitem[{{Ostorero} {et~al.}(2004){Ostorero}, {Villata}, \& {Raiteri}}]{ost04}
{Ostorero}, L., {Villata}, M., \& {Raiteri}, C.~M. 2004, \aap, 419, 913

\bibitem[{{O'Sullivan} \& {Gabuzda}(2009)}]{osu09}
{O'Sullivan}, S.~P. \& {Gabuzda}, D.~C. 2009, \mnras, 393, 429

\bibitem[{{Padovani} {et~al.}(2001){Padovani}, {Costamante}, {Giommi},
  {Ghisellini}, {Comastri}, {Wolter}, {Maraschi}, {Tagliaferri}, \& {Megan
  Urry}}]{pad01}
{Padovani}, P., {Costamante}, L., {Giommi}, P., {et~al.} 2001, \mnras, 328, 931

\bibitem[{{Papadakis} {et~al.}(2007){Papadakis}, {Villata}, \&
  {Raiteri}}]{pap07}
{Papadakis}, I.~E., {Villata}, M., \& {Raiteri}, C.~M. 2007, \aap, 470, 857

\bibitem[{{Pian} {et~al.}(1999){Pian}, {Urry}, {Maraschi}, {Madejski},
  {McHardy}, {Koratkar}, {Treves}, {Chiappetti}, {Grandi}, {Hartman}, {Kubo},
  {Leach}, {Pesce}, {Imhoff}, {Thompson}, \& {Wehrle}}]{pia99}
{Pian}, E., {Urry}, C.~M., {Maraschi}, L., {et~al.} 1999, \apj, 521, 112

\bibitem[{{Pucella} {et~al.}(2008){Pucella}, {Vittorini}, {D'Ammando},
  {Tavani}, {Raiteri}, {Villata}, {Argan}, {Barbiellini}, {Boffelli},
  {Bulgarelli}, {Caraveo}, {Cattaneo}, {Chen}, {Cocco}, {Costa}, {Del Monte},
  {de Paris}, {di Cocco}, {Donnarumma}, {Evangelista}, {Feroci}, {Fiorini},
  {Froysland}, {Fuschino}, {Galli}, {Gianotti}, {Giuliani}, {Labanti},
  {Lapshov}, {Lazzarotto}, {Lipari}, {Longo}, {Marisaldi}, {Mereghetti},
  {Morselli}, {Pacciani}, {Pellizzoni}, {Perotti}, {Picozza}, {Prest},
  {Rapisarda}, {Rappoldi}, {Soffitta}, {Trifoglio}, {Trois}, {Vallazza},
  {Vercellone}, {Zambra}, {Zanello}, {Antonelli}, {Colafrancesco}, {Cutini},
  {Gasparrini}, {Giommi}, {Pittori}, {Verrecchia}, {Salotti}, {Aller}, {Aller},
  {Carosati}, {Larionov}, \& {Ligustri}}]{puc08}
{Pucella}, G., {Vittorini}, V., {D'Ammando}, F., {et~al.} 2008, \aap, 491, L21

\bibitem[{{Raiteri} \& {Villata}(2003)}]{rai03p}
{Raiteri}, C.~M. \& {Villata}, M. 2003, in Proc.\ of the First ENIGMA Meeting,
  held at Mayschoss, Germany, May 11-14, 2003, Eds.: M.\ Hauser, U.\ Bach, \&
  S.\ Britzen, 326

\bibitem[{{Raiteri} {et~al.}(2007{\natexlab{a}}){Raiteri}, {Villata},
  {Capetti}, {Heidt}, {Arnaboldi}, \& {Magazz{\`u}}}]{rai07a}
{Raiteri}, C.~M., {Villata}, M., {Capetti}, A., {et~al.} 2007{\natexlab{a}},
  \aap, 464, 871

\bibitem[{{Raiteri} {et~al.}(2005){Raiteri}, {Villata}, {Ibrahimov},
  {Larionov}, {Kadler}, {Aller}, {Aller}, {Kovalev}, {Lanteri}, {Nilsson},
  {Papadakis}, {Pursimo}, {Romero}, {Ter{\"a}sranta}, {Tornikoski}, {Arkharov},
  {Barnaby}, {Berdyugin}, {B{\"o}ttcher}, {Byckling}, {Carini}, {Carosati},
  {Cellone}, {Ciprini}, {Combi}, {Crapanzano}, {Crowe}, {di Paola}, {Dolci},
  {Fuhrmann}, {Gu}, {Hagen-Thorn}, {Hakala}, {Impellizzeri}, {Jorstad}, {Kerp},
  {Kimeridze}, {Kovalev}, {Kraus}, {Krichbaum}, {Kurtanidze},
  {L{\"a}hteenm{\"a}ki}, {Lindfors}, {Mingaliev}, {Nesci}, {Nikolashvili},
  {Ohlert}, {Orio}, {Ostorero}, {Pasanen}, {Pati}, {Poteet}, {Ros}, {Ros},
  {Shastri}, {Sigua}, {Sillanp{\"a}{\"a}}, {Smith}, {Takalo}, {Tosti},
  {Vasileva}, {Wagner}, {Walters}, {Webb}, {Wills}, {Witzel}, \&
  {Xilouris}}]{rai05}
{Raiteri}, C.~M., {Villata}, M., {Ibrahimov}, M.~A., {et~al.} 2005, \aap, 438,
  39

\bibitem[{{Raiteri} {et~al.}(2006{\natexlab{a}}){Raiteri}, {Villata}, {Kadler},
  {Ibrahimov}, {Kurtanidze}, {Larionov}, {Tornikoski}, {Boltwood}, {Lee},
  {Aller}, {Romero}, {Aller}, {Araudo}, {Arkharov}, {Bach}, {Barnaby},
  {Berdyugin}, {Buemi}, {Carini}, {Carosati}, {Cellone}, {Cool}, {Dolci},
  {Efimova}, {Fuhrmann}, {Hagen-Thorn}, {Holcomb}, {Ilyin}, {Impellizzeri},
  {Ivanidze}, {Kapanadze}, {Kerp}, {Konstantinova}, {Kovalev}, {Kovalev},
  {Kraus}, {Krichbaum}, {L{\"a}hteenm{\"a}ki}, {Lanteri}, {Leto}, {Lindfors},
  {Mattox}, {Napoleone}, {Nikolashvili}, {Nilsson}, {Ohlert}, {Papadakis},
  {Pasanen}, {Poteet}, {Pursimo}, {Ros}, {Sigua}, {Smith}, {Takalo},
  {Trigilio}, {Tr{\"o}ller}, {Umana}, {Ungerechts}, {Walters}, {Witzel}, \&
  {Xilouris}}]{rai06b}
{Raiteri}, C.~M., {Villata}, M., {Kadler}, M., {et~al.} 2006{\natexlab{a}},
  \aap, 459, 731

\bibitem[{{Raiteri} {et~al.}(2006{\natexlab{b}}){Raiteri}, {Villata}, {Kadler},
  {Krichbaum}, {B{\"o}ttcher}, {Fuhrmann}, \& {Orio}}]{rai06a}
{Raiteri}, C.~M., {Villata}, M., {Kadler}, M., {et~al.} 2006{\natexlab{b}},
  \aap, 452, 845

\bibitem[{{Raiteri} {et~al.}(2008){Raiteri}, {Villata}, {Larionov}, {Gurwell},
  {Chen}, {Kurtanidze}, {Aller}, {B{\"o}ttcher}, {Calcidese}, {Hroch},
  {L{\"a}hteenm{\"a}ki}, {Lee}, {Nilsson}, {Ohlert}, {Papadakis}, {Agudo},
  {Aller}, {Angelakis}, {Arkharov}, {Bach}, {Bachev}, {Berdyugin}, {Buemi},
  {Carosati}, {Charlot}, {Chatzopoulos}, {Forn{\'e}}, {Frasca}, {Fuhrmann},
  {G{\'o}mez}, {Gupta}, {Hagen-Thorn}, {Hsiao}, {Jordan}, {Jorstad},
  {Konstantinova}, {Kopatskaya}, {Krichbaum}, {Lanteri}, {Larionova}, {Latev},
  {Le Campion}, {Leto}, {Lin}, {Marchili}, {Marilli}, {Marscher}, {McBreen},
  {Mihov}, {Nesci}, {Nicastro}, {Nikolashvili}, {Novak}, {Ovcharov}, {Pian},
  {Principe}, {Pursimo}, {Ragozzine}, {Ros}, {Sadun}, {Sagar}, {Semkov},
  {Smart}, {Smith}, {Strigachev}, {Takalo}, {Tavani}, {Tornikoski}, {Trigilio},
  {Uckert}, {Umana}, {Valcheva}, {Vercellone}, {Volvach}, \&
  {Wiesemeyer}}]{rai08c}
{Raiteri}, C.~M., {Villata}, M., {Larionov}, V.~M., {et~al.} 2008, \aap, 491,
  755

\bibitem[{{Raiteri} {et~al.}(2007{\natexlab{b}}){Raiteri}, {Villata},
  {Larionov}, {Pursimo}, {Ibrahimov}, {Nilsson}, {Aller}, {Kurtanidze},
  {Foschini}, {Ohlert}, {Papadakis}, {Sumitomo}, {Volvach}, {Aller},
  {Arkharov}, {Bach}, {Berdyugin}, {B{\"o}ttcher}, {Buemi}, {Calcidese},
  {Charlot}, {Delgado S{\'a}nchez}, {di Paola}, {Djupvik}, {Dolci}, {Efimova},
  {Fan}, {Forn{\'e}}, {Gomez}, {Gupta}, {Hagen-Thorn}, {Hooks}, {Hovatta},
  {Ishii}, {Kamada}, {Konstantinova}, {Kopatskaya}, {Kovalev}, {Kovalev},
  {L{\"a}hteenm{\"a}ki}, {Lanteri}, {Le Campion}, {Lee}, {Leto}, {Lin},
  {Lindfors}, {Mingaliev}, {Mizoguchi}, {Nicastro}, {Nikolashvili},
  {Nishiyama}, {{\"O}stman}, {Ovcharov}, {P{\"a}{\"a}kk{\"o}nen}, {Pasanen},
  {Pian}, {Rector}, {Ros}, {Sadakane}, {Selj}, {Semkov}, {Sharapov}, {Somero},
  {Stanev}, {Strigachev}, {Takalo}, {Tanaka}, {Tavani}, {Torniainen},
  {Tornikoski}, {Trigilio}, {Umana}, {Vercellone}, {Valcheva}, {Volvach}, \&
  {Yamanaka}}]{rai07b}
{Raiteri}, C.~M., {Villata}, M., {Larionov}, V.~M., {et~al.}
  2007{\natexlab{b}}, \aap, 473, 819

\bibitem[{{Raiteri} {et~al.}(1999){Raiteri}, {Villata}, {Tosti}, {Fiorucci},
  {Ghisellini}, {Takalo}, {Sillanp{\"a}{\"a}}, {Valtaoja}, {Ter{\"a}sranta},
  {Tornikoski}, {Aller}, {Aller}, {De Francesco}, {Hein{\"a}m{\"a}ki},
  {Katajainen}, {Lanteri}, {Nilsson}, {Pursimo}, {Rizzi}, \& {Sobrito}}]{rai99}
{Raiteri}, C.~M., {Villata}, M., {Tosti}, G., {et~al.} 1999, \aap, 352, 19

\bibitem[{{Ravasio} {et~al.}(2002){Ravasio}, {Tagliaferri}, {Ghisellini},
  {Giommi}, {Nesci}, {Massaro}, {Chiappetti}, {Celotti}, {Costamante},
  {Maraschi}, {Tavecchio}, {Tosti}, {Treves}, {Wolter}, {Balonek}, {Carini},
  {Kato}, {Kurtanidze}, {Montagni}, {Nikolashvili}, {Noble}, {Nucciarelli},
  {Raiteri}, {Sclavi}, {Uemura}, \& {Villata}}]{rav02}
{Ravasio}, M., {Tagliaferri}, G., {Ghisellini}, G., {et~al.} 2002, \aap, 383,
  763

\bibitem[{{Ravasio} {et~al.}(2003){Ravasio}, {Tagliaferri}, {Ghisellini},
  {Tavecchio}, {B{\"o}ttcher}, \& {Sikora}}]{rav03}
{Ravasio}, M., {Tagliaferri}, G., {Ghisellini}, G., {et~al.} 2003, \aap, 408,
  479

\bibitem[{{Sambruna} {et~al.}(1999){Sambruna}, {Ghisellini}, {Hooper},
  {Kollgaard}, {Pesce}, \& {Urry}}]{sam99}
{Sambruna}, R.~M., {Ghisellini}, G., {Hooper}, E., {et~al.} 1999, \apj, 515,
  140

\bibitem[{{Scarpa} {et~al.}(2000){Scarpa}, {Urry}, {Falomo}, {Pesce}, \&
  {Treves}}]{sca00}
{Scarpa}, R., {Urry}, C.~M., {Falomo}, R., {Pesce}, J.~E., \& {Treves}, A.
  2000, \apj, 532, 740

\bibitem[{{Schlegel} {et~al.}(1998){Schlegel}, {Finkbeiner}, \&
  {Davis}}]{sch98}
{Schlegel}, D.~J., {Finkbeiner}, D.~P., \& {Davis}, M. 1998, \apj, 500, 525

\bibitem[{{Shakura} \& {Sunyaev}(1973)}]{sha73}
{Shakura}, N.~I. \& {Sunyaev}, R.~A. 1973, \aap, 24, 337

\bibitem[{{Sikora} {et~al.}(1994){Sikora}, {Begelman}, \& {Rees}}]{sik94}
{Sikora}, M., {Begelman}, M.~C., \& {Rees}, M.~J. 1994, \apj, 421, 153

\bibitem[{{Smith} {et~al.}(1993){Smith}, {Schmidt}, \& {Allen}}]{smi93}
{Smith}, P.~S., {Schmidt}, G.~D., \& {Allen}, R.~G. 1993, \apj, 409, 604

\bibitem[{{Stickel} {et~al.}(1991){Stickel}, {Fried}, {Kuehr}, {Padovani}, \&
  {Urry}}]{sti91}
{Stickel}, M., {Fried}, J.~W., {Kuehr}, H., {Padovani}, P., \& {Urry}, C.~M.
  1991, \apj, 374, 431

\bibitem[{{Stirling} {et~al.}(2003){Stirling}, {Cawthorne}, {Stevens},
  {Jorstad}, {Marscher}, {Lister}, {G{\'o}mez}, {Smith}, {Agudo}, {Gabuzda},
  {Robson}, \& {Gear}}]{sti03}
{Stirling}, A.~M., {Cawthorne}, T.~V., {Stevens}, J.~A., {et~al.} 2003, \mnras,
  341, 405

\bibitem[{{Str{\"u}der} {et~al.}(2001){Str{\"u}der}, {Briel}, {Dennerl},
  {Hartmann}, {Kendziorra}, {Meidinger}, {Pfeffermann}, {Reppin}, {Aschenbach},
  {Bornemann}, {Br{\"a}uninger}, {Burkert}, {Elender}, {Freyberg}, {Haberl},
  {Hartner}, {Heuschmann}, {Hippmann}, {Kastelic}, {Kemmer}, {Kettenring},
  {Kink}, {Krause}, {M{\"u}ller}, {Oppitz}, {Pietsch}, {Popp}, {Predehl},
  {Read}, {Stephan}, {St{\"o}tter}, {Tr{\"u}mper}, {Holl}, {Kemmer}, {Soltau},
  {St{\"o}tter}, {Weber}, {Weichert}, {von Zanthier}, {Carathanassis}, {Lutz},
  {Richter}, {Solc}, {B{\"o}ttcher}, {Kuster}, {Staubert}, {Abbey}, {Holland},
  {Turner}, {Balasini}, {Bignami}, {La Palombara}, {Villa}, {Buttler},
  {Gianini}, {Lain{\'e}}, {Lumb}, \& {Dhez}}]{str01}
{Str{\"u}der}, L., {Briel}, U., {Dennerl}, K., {et~al.} 2001, \aap, 365, L18

\bibitem[{{Tanihata} {et~al.}(2000){Tanihata}, {Takahashi}, {Kataoka},
  {Madejski}, {Inoue}, {Kubo}, {Makino}, {Mattox}, \& {Kawai}}]{tan00}
{Tanihata}, C., {Takahashi}, T., {Kataoka}, J., {et~al.} 2000, \apj, 543, 124

\bibitem[{{Tateyama} {et~al.}(1998){Tateyama}, {Kingham}, {Kaufmann}, {Piner},
  {de Lucena}, \& {Botti}}]{tat98}
{Tateyama}, C.~E., {Kingham}, K.~A., {Kaufmann}, P., {et~al.} 1998, \apj, 500,
  810

\bibitem[{{T{\"u}rler} {et~al.}(2006){T{\"u}rler}, {Chernyakova},
  {Courvoisier}, {Foellmi}, {Aller}, {Aller}, {Kraus}, {Krichbaum},
  {L{\"a}hteenm{\"a}ki}, {Marscher}, {McHardy}, {O'Brien}, {Page}, {Popescu},
  {Robson}, {Tornikoski}, \& {Ungerechts}}]{tur06}
{T{\"u}rler}, M., {Chernyakova}, M., {Courvoisier}, T.~J.-L., {et~al.} 2006,
  \aap, 451, L1

\bibitem[{{Turner} {et~al.}(2001){Turner}, {Abbey}, {Arnaud}, {Balasini},
  {Barbera}, {Belsole}, {Bennie}, {Bernard}, {Bignami}, {Boer}, {Briel},
  {Butler}, {Cara}, {Chabaud}, {Cole}, {Collura}, {Conte}, {Cros}, {Denby},
  {Dhez}, {Di Coco}, {Dowson}, {Ferrando}, {Ghizzardi}, {Gianotti}, {Goodall},
  {Gretton}, {Griffiths}, {Hainaut}, {Hochedez}, {Holland}, {Jourdain},
  {Kendziorra}, {Lagostina}, {Laine}, {La Palombara}, {Lortholary}, {Lumb},
  {Marty}, {Molendi}, {Pigot}, {Poindron}, {Pounds}, {Reeves}, {Reppin},
  {Rothenflug}, {Salvetat}, {Sauvageot}, {Schmitt}, {Sembay}, {Short},
  {Spragg}, {Stephen}, {Str{\"u}der}, {Tiengo}, {Trifoglio}, {Tr{\"u}mper},
  {Vercellone}, {Vigroux}, {Villa}, {Ward}, {Whitehead}, \& {Zonca}}]{tur01}
{Turner}, M.~J.~L., {Abbey}, A., {Arnaud}, M., {et~al.} 2001, \aap, 365, L27

\bibitem[{{Vercellone} {et~al.}(2009){Vercellone}, {Chen}, {Vittorini},
  {Giuliani}, {D'Ammando}, {Tavani}, {Donnarumma}, {Pucella}, {Raiteri},
  {Villata}, {Chen}, {Tosti}, {Impiombato}, {Romano}, {Belfiore}, {DeLuca},
  {Novara}, {Senziani}, {Bazzano}, {Fiocchi}, {Ubertini}, {Ferrari}, {Argan},
  {Barbiellini}, {Boffelli}, {Bulgarelli}, {Caraveo}, {Cattaneo}, {Cocco},
  {Costa}, {Monte}, {DeParis}, {Cocco}, {Evangelista}, {Feroci}, {Fiorini},
  {Fornari}, {Froysland}, {Fuschino}, {Galli}, {Gianotti}, {Labanti},
  {Lapshov}, {Lazzarotto}, {Lipari}, {Longo}, {Marisaldi}, {Mereghetti},
  {Morselli}, {Pellizzoni}, {Pacciani}, {Perotti}, {Picozza}, {Prest},
  {Rapisarda}, {Rappoldi}, {Soffitta}, {Trifoglio}, {Trois}, {Vallazza},
  {Zambra}, {Zanello}, {Pittori}, {Verrecchia}, {Santolamazza}, {Preger},
  {Gasparrini}, {Cutini}, {Giommi}, {Colafrancesco}, \& {Salotti}}]{ver09}
{Vercellone}, S., {Chen}, A.~W., {Vittorini}, V., {et~al.} 2009, \apj, 690,
  1018

\bibitem[{{Vermeulen} {et~al.}(1995){Vermeulen}, {Ogle}, {Tran}, {Browne},
  {Cohen}, {Readhead}, {Taylor}, \& {Goodrich}}]{ver95}
{Vermeulen}, R.~C., {Ogle}, P.~M., {Tran}, H.~D., {et~al.} 1995, \apjl, 452,
  L5+

\bibitem[{{Villata} \& {Raiteri}(1999)}]{vil99}
{Villata}, M. \& {Raiteri}, C.~M. 1999, \aap, 347, 30

\bibitem[{{Villata} {et~al.}(2004{\natexlab{a}}){Villata}, {Raiteri}, {Aller},
  {Aller}, {Ter{\"a}sranta}, {Koivula}, {Wiren}, {Kurtanidze}, {Nikolashvili},
  {Ibrahimov}, {Papadakis}, {Tosti}, {Hroch}, {Takalo}, {Sillanp{\"a}{\"a}},
  {Hagen-Thorn}, {Larionov}, {Schwartz}, {Basler}, {Brown}, \&
  {Balonek}}]{vil04b}
{Villata}, M., {Raiteri}, C.~M., {Aller}, H.~D., {et~al.} 2004{\natexlab{a}},
  \aap, 424, 497

\bibitem[{{Villata} {et~al.}(2004{\natexlab{b}}){Villata}, {Raiteri},
  {Kurtanidze}, {Nikolashvili}, {Ibrahimov}, {Papadakis}, {Tosti}, {Hroch},
  {Takalo}, {Sillanp{\"a}{\"a}}, {Hagen-Thorn}, {Larionov}, {Schwartz},
  {Basler}, {Brown}, {Balonek}, {Ben{\'{\i}}tez}, {Ram{\'{\i}}rez}, {Sadun},
  {Boltwood}, {Carini}, {Barnaby}, {Coloma}, {Ros}, {Dai}, {Xie}, {Mattox},
  {Rodriguez}, {Asfandiyarov}, {Atkerson}, {Beem}, {Bloom}, {Chanturiya},
  {Ciprini}, {Crapanzano}, {de Diego}, {Efimova}, {Gardiol}, {Guerra},
  {Kahharov}, {Kapanadze}, {Karttunen}, {Kato}, {Kimeridze}, {Kudryavtseva},
  {Lainela}, {Lanteri}, {Larionova}, {Maesano}, {Marchili}, {Massone},
  {Monroe}, {Montagni}, {Nesci}, {Nilsson}, {Noble}, {Nucciarelli}, {Ostorero},
  {Papamastorakis}, {Pasanen}, {Peters}, {Pursimo}, {Reig}, {Ryle}, {Sclavi},
  {Sigua}, {Uemura}, \& {Wills}}]{vil04a}
{Villata}, M., {Raiteri}, C.~M., {Kurtanidze}, O.~M., {et~al.}
  2004{\natexlab{b}}, \aap, 421, 103

\bibitem[{{Villata} {et~al.}(2002){Villata}, {Raiteri}, {Kurtanidze},
  {Nikolashvili}, {Ibrahimov}, {Papadakis}, {Tsinganos}, {Sadakane}, {Okada},
  {Takalo}, {Sillanp{\"a}{\"a}}, {Tosti}, {Ciprini}, {Frasca}, {Marilli},
  {Robb}, {Noble}, {Jorstad}, {Hagen-Thorn}, {Larionov}, {Nesci}, {Maesano},
  {Schwartz}, {Basler}, {Gorham}, {Iwamatsu}, {Kato}, {Pullen},
  {Ben{\'{\i}}tez}, {de Diego}, {Moilanen}, {Oksanen}, {Rodriguez}, {Sadun},
  {Kelly}, {Carini}, {Miller}, {Catalano}, {Dultzin-Hacyan}, {Fan}, {Ishioka},
  {Karttunen}, {Kein{\"a}nen}, {Kudryavtseva}, {Lainela}, {Lanteri},
  {Larionova}, {Matsumoto}, {Mattox}, {Montagni}, {Nucciarelli}, {Ostorero},
  {Papamastorakis}, {Pasanen}, {Sobrito}, \& {Uemura}}]{vil02}
{Villata}, M., {Raiteri}, C.~M., {Kurtanidze}, O.~M., {et~al.} 2002, \aap, 390,
  407

\bibitem[{{Villata} {et~al.}(2009){Villata}, {Raiteri}, {Larionov},
  {Nikolashvili}, {Aller}, {Bach}, {Carosati}, {Hroch}, {Ibrahimov}, {Jorstad},
  {Kovalev}, {L{\"a}hteenm{\"a}ki}, {Nilsson}, {Ter{\"a}sranta}, {Tosti},
  {Aller}, {Arkharov}, {Berdyugin}, {Boltwood}, {Buemi}, {Casas}, {Charlot},
  {Coloma}, {di Paola}, {di Rico}, {Kimeridze}, {Konstantinova}, {Kopatskaya},
  {Kovalev}, {Kurtanidze}, {Lanteri}, {Larionova}, {Larionova}, {Le Campion},
  {Leto}, {Lindfors}, {Marscher}, {Marshall}, {McFarland}, {McHardy}, {Miller},
  {Nucciarelli}, {Osterman}, {Pasanen}, {Pursimo}, {Ros}, {Sadun}, {Sigua},
  {Sixtova}, {Takalo}, {Tornikoski}, {Trigilio}, {Umana}, {Xie}, {Zhang}, \&
  {Zhou}}]{vil09}
{Villata}, M., {Raiteri}, C.~M., {Larionov}, V.~M., {et~al.} 2009, \aap, 501,
  455

\bibitem[{{von Montigny} {et~al.}(1997){von Montigny}, {Aller}, {Aller},
  {Bruhweiler}, {Collmar}, {Courvoisier}, {Edwards}, {Fichtel}, {Fruscione},
  {Ghisellini}, {Hartman}, {Johnson}, {Kafatos}, {Kii}, {Kniffen}, {Lichti},
  {Makino}, {Mannheim}, {Marscher}, {McBreen}, {McHardy}, {Pesce}, {Pohl},
  {Ramos}, {Reich}, {Robson}, {Sasaki}, {Teraesranta}, {Tornikoski}, {Urry},
  {Valtaoja}, {Wagner}, \& {Weekes}}]{von97}
{von Montigny}, C., {Aller}, H., {Aller}, M., {et~al.} 1997, \apj, 483, 161

\bibitem[{{Wilms} {et~al.}(2000){Wilms}, {Allen}, \& {McCray}}]{wil00}
{Wilms}, J., {Allen}, A., \& {McCray}, R. 2000, \apj, 542, 914

\end{thebibliography}
\end{document}